\documentclass[11pt,a4paper]{article}
\pdfoutput=1
\usepackage{jcappub}

\usepackage{graphicx}
\usepackage{dcolumn}
\usepackage{color}
\usepackage{bm}
\usepackage{hyperref}
\usepackage[mathlines]{lineno}
\usepackage{amsfonts,amsmath,amssymb,bbm}
\usepackage{titlesec}
\usepackage{verbatim}
\usepackage{fontawesome}

\title{Implications of a transition in the dark energy equation of state for the $H_0$ and $\sigma_8$ tensions}

\author[a,b,1]{Ryan E.~Keeley, \note{Corresponding author.}}
\author[c]{Shahab Joudaki,}
\author[b]{Manoj Kaplinghat,}
\author[b]{David Kirkby}

\affiliation[a]{Korea Astronomy and Space Science Institute, Yuseong-gu, 776 Daedeok-daero, Daejeon, Korea}
\affiliation[b]{Center for Cosmology, Department of Physics and Astronomy, University of California, Irvine, CA 92697, USA}
\affiliation[c]{Department of Physics, University of Oxford, Denys Wilkinson Building, Keble Road, Oxford OX1 3RH, U.K.}

\emailAdd{rkeeley@kasi.re.kr}

\abstract{
We explore the implications of a rapid appearance of dark energy between the redshifts ($z$) of one and two on the expansion rate and growth of perturbations.
Using both Gaussian process regression and a parametric model, we show that this is the preferred solution to the current set of low-redshift ($z<3$) distance measurements if $H_0=73~\rm km\,s^{-1}\,Mpc^{-1}$ to within 1\% and the high-redshift expansion history is unchanged from the $\Lambda$CDM inference by the Planck satellite.
Dark energy was effectively non-existent around $z=2$, but its density is close to the $\Lambda$CDM model value today, with an equation of state greater than $-1$ at $z<0.5$. If sources of clustering other than matter are negligible, we show that this expansion history leads to slower growth of perturbations at $z<1$, compared to $\Lambda$CDM, that is measurable by upcoming surveys and can alleviate the $\sigma_8$ tension between the Planck CMB temperature and low-redshift probes of the large-scale structure.
}

\arxivnumber{1905.10198}

\begin{document}
\maketitle
\flushbottom

\section{The tensions}
\label{sec:Intro}

The current concordance cosmological model consisting of a cosmological constant ($\Lambda$) and cold dark matter (CDM) has been remarkably successful in explaining cosmological observables at both high and low redshift \citep{planck15,alam16,Betoule:2014frx, Hildebrandt16,Aghanim:2018eyx}.  However, within this $\Lambda$CDM model, some tensions between datasets have emerged that merit attention. One is the ``$H_0$ tension'', which is a mismatch between the direct measurement of the present expansion rate, or Hubble constant, and the value inferred from observations of the cosmic microwave background (CMB)~\cite{Riess:2016jrr,riess19,Aghanim:2018eyx}.

The second is the ``$\sigma_8$ tension'', which is a discrepancy between the RMS of the linear matter density field ($\sigma(R,z)$) on $8 \, h^{-1} \, {\rm Mpc}$ scales at redshift $z=0$ ($\sigma_8$) inferred from the CMB and measured by Sunyaev-Zel'dovich (SZ) cluster counts (e.g.~\citep{sz2014,sz2016,Bocquet:2018ukq}). It should be noted that this tension depends on the adopted calibration of the SZ flux to cluster halo mass \cite{Pan:2018zha}, which is still uncertain. There are also indications that the value of $S_8=\sigma_8 \Omega_{\rm m}^{0.5}$, where $\Omega_{\rm m}$ is the matter density relative to the critical density today, as measured through weak gravitational lensing tomography is in tension with the inference from the CMB (e.g.~\cite{heymans13,joudaki16a,Hildebrandt16,joudaki16b,Kohlinger:2017sxk,Joudaki:2017zdt,vanUitert:2017ieu,Troxel:2017xyo,Abbott:2017wau,Hikage:2018qbn}).

In addition to the discordance between the direct measurement of $H_0$ and the CMB, there is also a discordance between the distances calibrated by them.  The supernova (SNe) distances calibrated by the local $H_0$ measurement do not agree at $z\simeq 0.5$ with the  distances inferred from the baryon acoustic oscillation (BAO) feature in the correlation function of luminous red galaxies (LRGs)~\cite{Beutler:2016ixs} calibrated by the CMB~\cite{Aylor:2018drw}.
One way to bring the SNe and LRG distances into agreement is if the true $r_{\rm drag}$ is smaller than the value from the $\Lambda$CDM fit to the Planck data~\cite{Aylor:2018drw}.

Though these disagreements could be due to unknown systematic uncertainties in the measurements, an interesting possibility is that these tensions point to new physics.  This point of view has merit because $H_0$ and $\sigma_8$ obtained from the CMB are derived parameters calculated from a model-dependent projection over three orders of magnitude in the scale factor, $a$. One way to evolve the Universe to low redshift in a manner different from $\Lambda$CDM is to relax the requirement that the dark energy is a cosmological constant. Evolving dark energy is often quantified by the Chevallier-Polarski-Linder parametrization \cite{Chevallier:2000qy,Linder:2002et}, but more dramatic changes are possible.
For instance, scalar-tensor theories satisfying the LIGO/Virgo bound on the propagation speed of gravitational waves \cite{Monitor:2017mdv} can include a dark energy component with equation of state that varies rapidly with redshift \cite{Kase:2018iwp,DeFelice:2010pv}.

An alternative possibility is that new physics at or before last scattering gives rise to a larger expansion rate in the early Universe and smaller sound horizon at the drag epoch, $r_{\rm drag}$, compared to $\Lambda$CDM. This shifts the CMB prediction and the low-redshift (BAO) features to better agree with the measured value of $H_0$ \cite{Bernal:2016gxb,Aylor:2018drw,Poulin:2018cxd}. The difficulty in this strategy comes from the fact that the CMB anisotropies are precisely measured~\cite{Aghanim:2018eyx}. It is difficult to modify the distance scales away from $\Lambda$CDM predictions without spoiling its successful predictions for the temperature and polarization anisotropies \cite{Bernal:2016gxb}.

Models with new physics at $z>1000$ typically add one or more extra degrees of freedom. Three possibilities that have been studied are dark radiation \cite{Bernal:2016gxb}, strongly interacting massive neutrinos \cite{Kreisch:2019yzn}, and early dark energy \cite{Poulin:2018cxd}. By adding extra degrees of freedom, any predictions for low-redshift quantities from these models will be more uncertain relative to the $\Lambda$CDM prediction, and currently proposed modifications at $z>1000$ use this reduction in significance to alleviate the $H_0$ tension  \cite{Joudaki:2017zhq,Bernal:2016gxb,Poulin:2018cxd,Kreisch:2019yzn}. However, should either the CMB polarization power spectrum or $H_0$ measurements become more precise while maintaining the same central values, the tension would reemerge.

An appealing aspect of these high-$z$ modifications that lead to smaller $r_{\rm drag}$ is that it can make $z<1$ distance measurements consistent with each other~\cite{Aylor:2018drw}. Our work in this paper asks the complementary question of how these tensions may be alleviated if the expansion rate at last scattering and $r_{\rm drag}$ are unchanged from the $\Lambda$CDM inferences. In this case, the path forward is to explore the uncertainty in the dark energy density evolution. We do so in a model-independent (Gaussian process (GP) regression) framework, and with a parametric model (Transitional Dark Energy (TDE) model) for dark energy evolution. Our approach highlights the interesting possibility of evolving dark energy models that alleviate the $H_0$ tension and predict a slower growth of structure at late times~\cite{joudaki16b}.

\section{Model-independent expansion history}\label{sec:GP}
We use a GP regression to infer the expansion history of the Universe following the procedure used in our previous work~\cite[][hereafter J18]{Joudaki:2017zhq}. The repository for that code is archived here~\cite{gphistdoi}. The present analysis differs from J18 in that we forecast results assuming a precision measurement of the Hubble constant at the 1\% level, a remarkable feat that could be achieved in the near future.  This may be possible through better calibration of Cepheids with Gaia  \cite{Freedman:2017yms,2012arXiv1202.4459S,Riess:2016jrr,2018ApJ...855..136R}, through a standard siren technique with LIGO and Virgo \cite{Chen:2017rfc}, via Population~II distance indicators with Gaia \cite{Beaton:2016nsw}, or via strong lensing time-delay measurements by the H0LiCOW collaboration \cite{Birrer:2018vtm}.

As in J18, we condition the GP regression using direct measurements of the Hubble distance $D_H(z) = c/H(z)$, as well as indirect constraints on $D_H$ from measurements of the angular diameter distance $D_A(z) = D_C(z) / (1+z)$, and the luminosity distance $D_L(z) = (1+z)D_C(z)$, where $D_C(z) = \int_0^z D_H(z) dz$. Throughout this work, we assume spatial flatness. We divide out a fiducial expansion history, $D_H^0$, based on the best-fit Planck+WP flat $\Lambda$CDM cosmology from Ade~et~al.~(2013)~\cite{planck13} (the differences between the 2013 and 2018 Planck results are small, and do not noticeably change the GP regression results) with $H_0 = 67.04~{\rm km}~{\rm s}^{-1}~{\rm Mpc}^{-1}$, present matter density $\Omega_{\rm m} = 0.3183$, present dark energy density $\Omega_{\rm DE} = 0.6817$, effective number of neutrinos $N_{\rm eff} = 3.046$, and one massive neutrino species with mass $m_\nu = 0.06$ eV.

We define a GP for $\gamma(z)=\ln(D_H(z)/D_H^0(z))$ with zero mean and covariance function
\begin{equation}
\langle \gamma(z_1)\gamma(z_2) \rangle = h^2 \exp(-(s(z_1) - s(z_2))^2/(2\sigma^2)),
\end{equation}
where the evolution variable $s(z)$ is taken to be $s(z) = \ln(1+z)/\ln(1+z_*)$,
and $z_* = 1090.48$ to match the redshift of last scattering for the Planck+WP best fit. Note that $s(z)$ goes from 0 to 1 as $z$ changes from 0 to $z_*$. We marginalize over the grid of hyperparameters $\{0.01<h<0.2,0.001<\sigma < 1.0 \}$.

\begin{figure*}[ht]
\centering
\includegraphics[width=\textwidth]{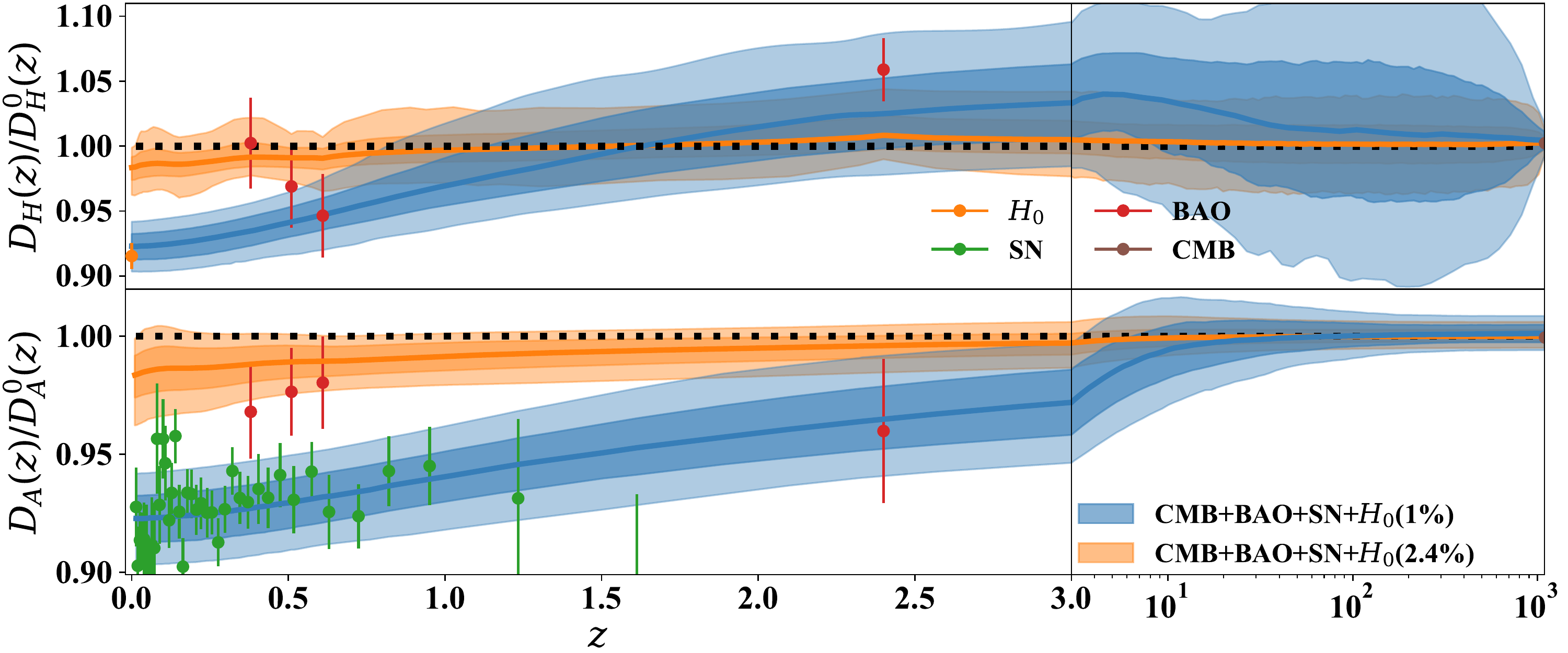}
\caption{Posteriors for the expansion history as determined by the GP regression (inner 68\% and outer 95\% confidence levels). The Hubble and angular diameter distances, $D_H(z)$ and $D_A(z)$, are shown in the top and bottom panels, respectively. These distances are shown relative to the fiducial Planck $\Lambda$CDM model. The orange shaded regions correspond to the results with the Riess~et~al.~(2016)~\cite{Riess:2016jrr} uncertainty on $H_0$ as calculated in J18, while the blue shaded regions correspond to forecasts with 1\% precision on $H_0$. The orange and blue solid lines illustrate the median results of the GPs.  Note the split linear-logarithmic redshift axis.
}
\label{fig:main}
\end{figure*}

We use the following datasets to constrain the GP:
\begin{itemize}
    \item Planck 2015 CMB temperature and polarization dataset consisting of `TT', `TE', `EE', and `lowP' angular power spectra \citep{planckone2015,planck15} which was used to compute the posterior mean and covariance for $D_H$ and $D_A$ at the redshift of last scattering, $z_*$.
    \item Distances inferred from the BAO signal encoded in the clustering of LRGs from Beutler~et~al.~(2016) \cite{Beutler:2016ixs}.
    \item Distances inferred from the BAO signal in the auto-correlation of the flux transmission of the Ly$\alpha$ forest and cross-correlation with quasars from Bourboux~et~al.~(2017)~\cite{Bourboux:2017cbm}. Each of these BAO distances scale with $r_{\rm drag}$, which we fix to the best-fit value from Planck 2015~\cite{planck15}.
    \item The `Pantheon' binned Type Ia SNe from Scolnic~et~al.~(2018)~\cite{Scolnic:2017caz}, which measure the ratio $D_L/D_{H_0}$.
    \item A direct Hubble constant measurement, similar to the 2.4\% determination from Riess~et~al.~(2016)~\cite{Riess:2016jrr}.
\end{itemize}

In this paper, we forecast results using the same posterior mean for $H_0$ as in Riess~et~al.~(2016)~\cite{Riess:2016jrr}, but with uncertainties of only 1\%. The updated GP regression code is archived here~\cite{gphistdoi2} and can be found in the public repository: \href{https://github.com/rekeeley/GPHistTDE}{\faGithub}\footnote{\url{https://github.com/rekeeley/GPHistTDE}}.
The recently released $H_0$ constraint in Riess~et~al.~(2019)~\cite{riess19}, after the completion of this work, is now at the level of 1.9\% and continues to be discrepant with the CMB-inferred value (now at $4.4\sigma$), further strengthening the approach taken here and conclusions reached in this paper.

Following J18, we infer the evolution of the dark energy ($\rho_{\rm DE}(z)$) and matter densities from the expansion rate, by assuming flatness and no new physics at last scattering. We also infer the dark energy equation of state through the energy conservation equation as $w(z)=-1-\rho_{\rm DE}'/(3\rho_{\rm DE})$, where the prime denotes derivative with respect to $\ln(a)$.

The results of the forecast GP regression appear in Fig.~\ref{fig:main}, which shows that the median inference (blue) favors $D_H$ values larger than the fiducial model for redshifts above $z\sim1.5$ and smaller below. At $z=0$, $D_A$ begins significantly below the fiducial values, eventually meeting them at $z=z_*$. Such a transition in $D_H$ arises from the need to satisfy the constraint on $D_A(z_*)$.
In Fig.~\ref{fig:main}, we also show the constraints from J18 (orange) using the current 2.4\% precision on $H_0$.  The error bars on the GP inference are smaller at high redshift (where there are no constraints) despite the increased precision in the $H_0$ measurement.  This is a feature of using GP as a prior. In GP regression the size of the error bars are tied to the size of the fluctuations away from the mean function (the fiducial cosmology) unless there is data to constrain it. As the increased precision on $H_0$ favors a significant deviation away from the fiducial cosmology, the error bars are larger at high redshift.

\begin{figure}
    \centering
    \includegraphics[width=0.45\textwidth]{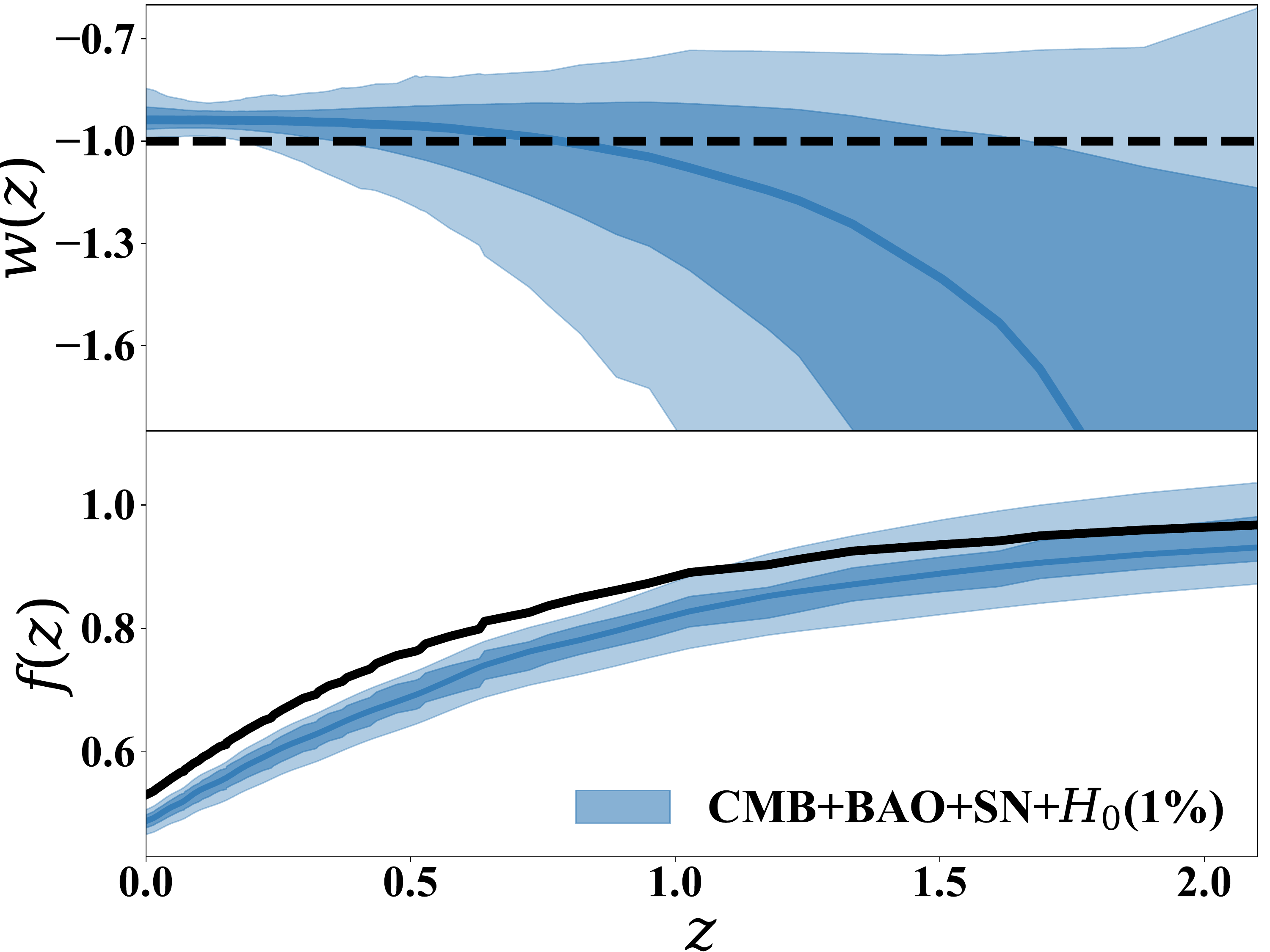}
    \caption{The top panel of the figure shows the inferred dark energy equation of state as a function of redshift from the GP regression. The bottom panel shows the growth rate $f = d\ln(D)/d\ln(a)$ from the GP regression.  As in Fig.~\ref{fig:main}, the blue shaded regions correspond to the 68\% and 95\% confidence levels and the solid line corresponds to the median of the GP inference. The black solid line in the bottom panel corresponds to the $\Lambda$CDM growth rate.
    }
    \label{fig:GP_w_f}
\end{figure}

The forecast inference of the dark energy evolution is shown in the upper panel of Fig.~\ref{fig:GP_w_f} and shows a transition in $w(z)$ that corresponds to the one in $D_H(z)$.
Here, the GP regression picks out a median value for $w(0)$ greater than $-1$ and, interestingly, the median inference quickly transitions to values much less than $-1$. This $w(z)$ behavior is consistent with that found necessary to reconcile current cosmic shear and Planck CMB temperature measurements \cite{joudaki16b}.

To understand why such an evolution in the dark energy component is preferred by the forecast data, note that the physical matter density at $z=0$ is set by the constraint on $D_H(z_*)$ (see  J18 for discussion). With this information known, the physical dark energy density at $z=0$ is then set by the large value of $H_0$. In the case of a cosmological constant, the inferred matter and dark energy density would make $D_H(z)$ too small to explain the observed value of $D_A(z_*)$.  Thus, in some interval between redshift zero and $z_*$, $D_H$ needs to be increased, and this can be only achieved by allowing the dark energy component to evolve. The other datasets constrain the redshift dependence of dark energy. For example, the SNe constrain the shape of $D_H(z)$ and hence how fast the dark energy can evolve at low redshifts, which explains why the evolution starts above $z=1$.

\begin{figure*}
    \centering
    \includegraphics[width=0.24\textwidth]{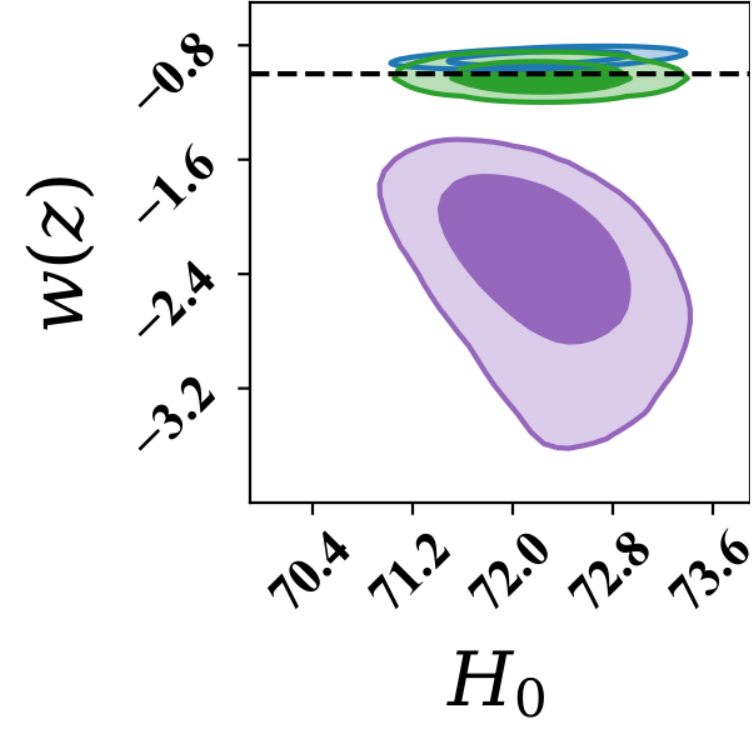}
    \includegraphics[width=0.24\textwidth]{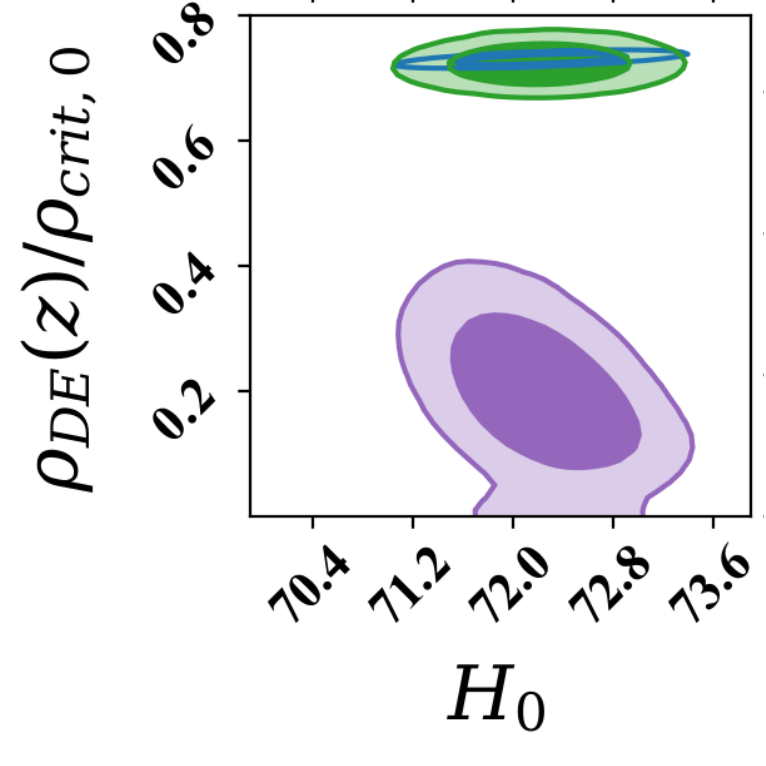}
    \includegraphics[width=0.24\textwidth]{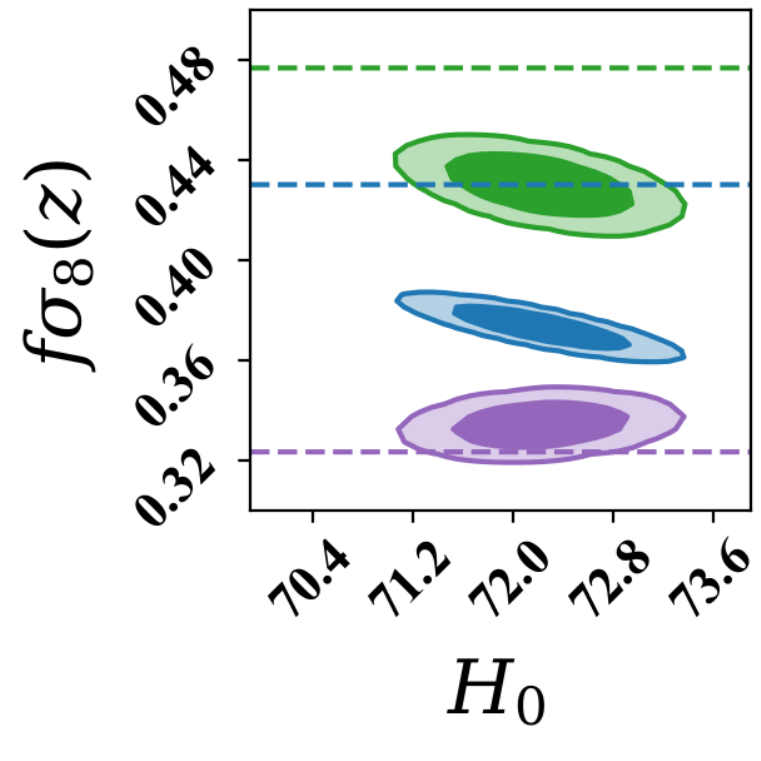}
    \includegraphics[width=0.24\textwidth]{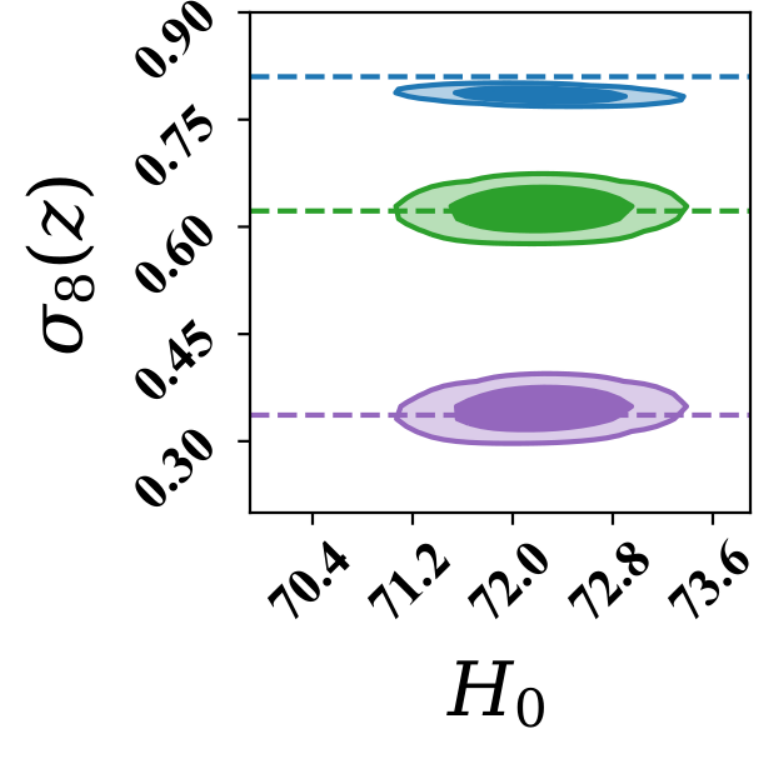}
    \caption{Posteriors of $H_0$ and representative derived parameters from the MCMC inference of the TDE model (inner 68\% CL, outer 95\% CL), fitting the same datasets as the GP. Each of these panels show the derived parameter evaluated at $z=0$, $z=0.5$, and $z=2$ (blue, green, violet). These representative parameters are the equation of state (left), dark energy density scaled to the present critical density (center left), $f\sigma_8$ (center right), and $\sigma_8$ (right). The dashed horizontal lines correspond to the fiducial $\Lambda$CDM values for $w(z)$ $f \sigma_8 (z)$ and $\sigma_8 (z)$.
    }
    \label{fig:MCMC_H0_de_growth}
\end{figure*}

The growth history is inferred in the same manner as in J18, by solving the following differential equation:
\begin{equation}\label{eq:growth}
\phi'' + (4+H'/H)\phi' + (3+2H'/H)\phi = 0 ,
\end{equation}
where $\phi$ is the gravitational potential. This equation can be derived from the spatial part of the perturbed Einstein equations in the conformal Newtonian gauge~\cite{mb95} by setting $\delta T^i_j=0$, i.e., no shear or pressure perturbations. Another way to derive this equation is to use the covariant conservation of the energy momentum tensor~\cite{mb95}, with the Poisson equation for $\phi$ on sub-horizon scales, and setting the two metric potentials ($\phi$ in the space-space part and $\psi$ in the time-time part of the metric) equal to each other and pressure perturbations to zero.

For a cosmology with pressure-less matter and a cosmological constant, Eq.~\ref{eq:growth} is the same as the usual growth equation $\ddot{\delta}_m+2H\dot{\delta}_m-4\pi G \rho_m \delta_m=0$ on sub-horizon scales, where $\rho_m \delta_m$ is the perturbation to the matter density and overdot denotes derivative with respect to coordinate time. Eq.~\ref{eq:growth} is a good way to explore modifications of the expansion history because early on ($z \gtrsim 2$) data prefers a matter-dominated cosmology and at late times ($z \lesssim 0.5$) data prefers dark energy with $w\simeq -1$. In making this assessment, we are implicitly assuming that the effective Gravitational constant appearing in the Poisson equation for the metric potential $\psi$ is the same as the Newtonian one and that the gravitational ``slip" \cite{Bertschinger:2006aw,Caldwell:2007cw} is negligible on small scales (i.e., $\phi/\psi = 1$). Our results later indicate that a gravitational slip is not required to match the observed growth history.

We define the growth function $D=a\phi$ and the growth rate $f = D'/D$. In Fig.~\ref{fig:GP_w_f}, we show how the inferred expansion history, which is significantly different from the fiducial $\Lambda$CDM expansion history, causes the corresponding growth history to differ from the fiducial $\Lambda$CDM expectation. The key result to note is that for $z<1$, the expansion rate $H(z)$ is larger than the fiducial expansion rate and hence the growth of perturbations is slower. This demonstrates that the $H_0$ and $\sigma_8$ tensions could have a common origin.

We have not provided a concrete model for the preferred dark energy evolution and the growth rate encapsulated in Eq.~\ref{eq:growth}. An interesting avenue to pursue is extensions to General Relativity (GR) that can motivate the kind of dark energy evolution that we have inferred. One such example is a new ``Galileon'' degree of freedom~\cite{DeFelice:2010pv}, which under the right initial conditions could have a dark energy equation of state $w=-2$ at high redshift and evolve towards $w=-1$ at low redshift, due to a de Sitter fixed point~\cite{Kase:2018iwp}.  Such a $w(z)$ evolution is broadly consistent with our results but there is more growth than predicted by Eq.~\ref{eq:growth}. There are also Generalized Proca theories (vector-tensor theories) with three propagating degrees of freedom where the early universe $w(z)$ could be $-1-s$ with a late Universe de Sitter attractor~\cite{Heisenberg:2014rta,DeFelice:2016yws}. However, $s$ is constrained from cosmological data (expansion history and growth) to be around $0.2$ \cite{deFelice:2017paw} and hence not consistent with solving the $H_0$ tension.

\section{Transitional Dark Energy model}\label{sec:Tanh}

\begin{figure}[ht]
\centering
\includegraphics[width=0.45\textwidth]{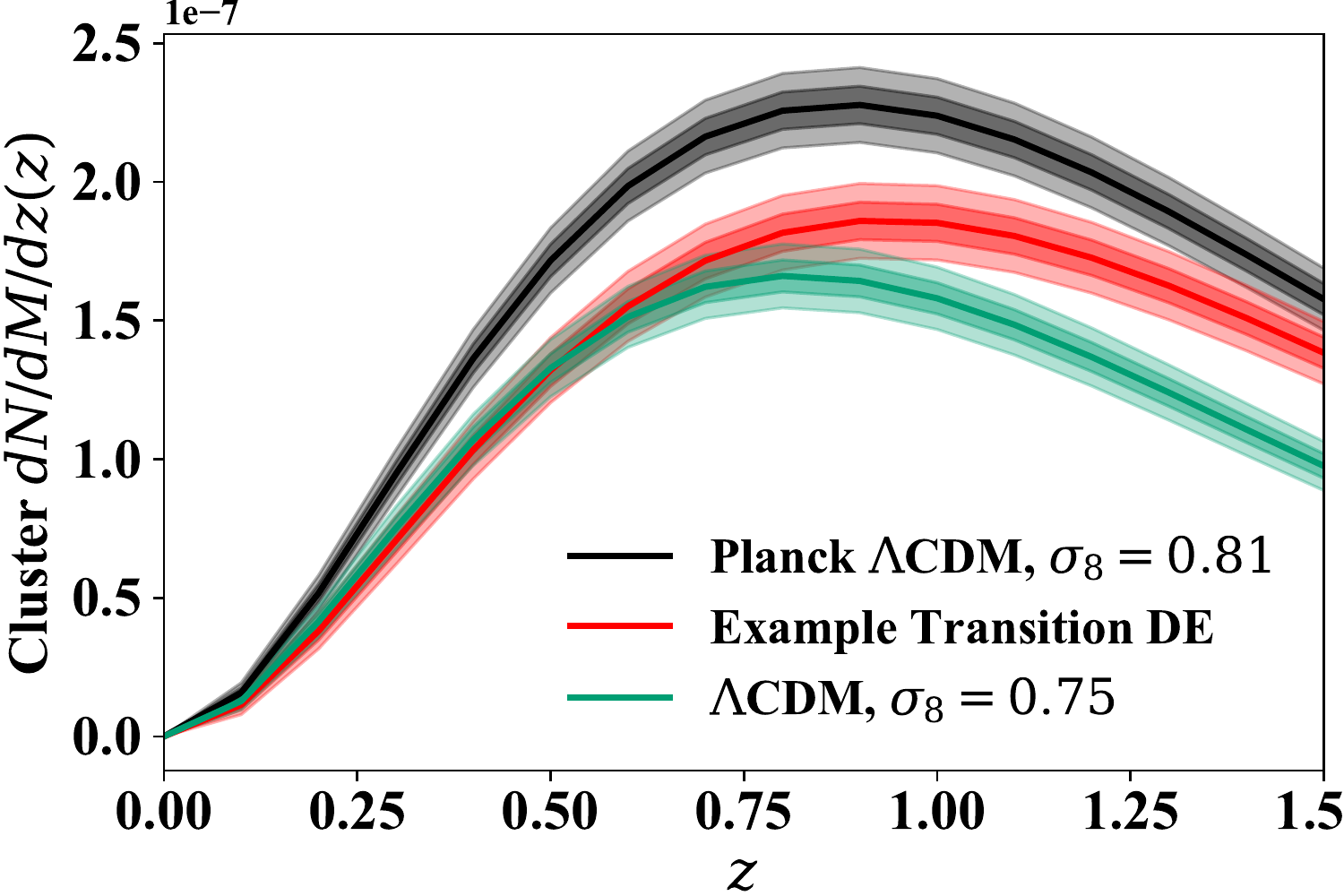}
\caption{Sunyaev-Zel'dovich cluster counts $dN/dM/dz$ for a $\Lambda$CDM model consistent with Planck (black), a $\Lambda$CDM model with $\sigma_8=0.75$ (green), and an example TDE model (red). The shaded bands correspond to the cosmic variance 68\% and 95\% CLs.
}
\label{fig:tanh}
\end{figure}

To further investigate the implications of a rapid change in the dark energy density, we switch to a concrete parameterization of the dark energy evolution. This allows us to compute observables that are sensitive to the growth history, such as SZ cluster counts \cite{Ade:2015fva} using the cosmological Boltzmann solver  \texttt{CLASS}~\cite{2011JCAP...07..034B}, which was modified to allow for a rapid transition in the dark energy equation of state.

To this end, we define the TDE model as,
\begin{eqnarray}
&&\rho_{\rm DE}(z) =\rho_{\rm DE,0}(1+z)^{3(1+W(z))},\\
    && W(z) = ((w_0+w_1) + (w_1-w_0)\tanh((z-z_t)/\Delta_z))/2,\nonumber
\end{eqnarray}
where $W(z)$ is related to the equation of state,
\begin{equation}
    W(z) = \frac{1}{\ln(1+z)} \int_{1/(1+z)}^1 w(a) \frac{da}{a}.
\end{equation}
This function is equivalent to the equation of state $w(z)$ in the regimes where $w(z)$ is constant.  The equation of state tends towards $w_0$ at $z<z_t$ and towards $w_1$ at $z>z_t$.  The width of this transition is parametrized by $\Delta_z$.  The values that fit the {\em median} GP inference are $w_0=-0.95$, $w_1=-1.95$, $z_t=2.5$, and $\Delta_z=0.9$. These values are used to calculate the growth functions in \texttt{CLASS}. Effectively, at early times the dark energy component is completely absent and then rapidly turns on by around redshift of 1. Similar models have been explored in the past~\cite{Bassett:2002qu,Shafieloo:2009ti}.

We performed a Markov Chain Monte Carlo (MCMC) analysis to sample the forecast posterior for the TDE model. We used the same datasets as in the forecast GP analysis, and varied all of the parameters of the TDE model ($h$, $\omega_{\rm m}$, $w_0$, $w_1$, $\Delta_z$, $z_T$) in the MCMC. In Fig.~\ref{fig:MCMC_H0_de_growth}, we show the forecast posteriors of the derived parameters $w(z)$, $\rho_{\rm DE}(z)/\rho_{\rm crit,0}$, $\sigma_8(z)$, and $f\sigma_8 (z)$ for redshifts $z=0,0.5,2$. These forecast posteriors illustrate that the datasets considered indeed would favor a drastic change in the dark energy equation of state at intermediate redshift, with little to no dark energy at $z=2$ and a relatively sharp transition in the redshift range 0.5 to 2.
We compute the Bayes factor ($K$) between the TDE model and the $\Lambda$CDM model and find $\ln K = 6.8$ in favor of the TDE model, corresponding to odds of 900 to 1 and `decisive' preference for the TDE model when using Jeffreys' scale~\cite{KassRaftery,Jeffreys}. A corresponding preference is found when computing the Deviance information criterion (DIC~\cite{spiegelhalter02}), with $\Delta \rm{DIC} = 24.2$.

The correlations shown in the forecast posteriors are particularly interesting. The fact that large values of $H_0$ are correlated with smaller amounts of dark energy (or equivalently, more negative values of the equation of state) at $z=2$ favors the idea that in order explain the observed value of $H_0 = 73 \,\rm km\,s^{-1}\,Mpc^{-1}$, along with all of the other considered datasets, dark energy must be evolving in some form \cite{DiValentino:2016hlg,joudaki16b,Zhao:2017cud}. Moreover, as larger values of $H_0$ are correlated with less growth at $z<2$ (most notably in $f\sigma_8$), this resolution to the $H_0$ problem would have interesting consequences for the $\sigma_8$ tension.
When using the 2.4\% uncertainty on $H_0$ from Riess~et~al.~(2016)~\cite{Riess:2016jrr} instead of the projected 1\% uncertainty, the posteriors for the TDE model become consistent with a cosmological constant (as in the GP regression from J18) and the global fit favors a value of $H_0$ around 69~km\,s$^{-1}$\,Mpc$^{-1}$.

In Sec.~\ref{app:CMBTDE} of the Appendix, we check that the TDE model can reproduce the best-fit $\Lambda$CDM $C_\ell$s by plotting the best-fit TDE parameters from a fit to the full Planck 2018 likelihoods~\cite{Aghanim:2018eyx}. Fig.~\ref{fig:Cell} shows the $C_\ell$s from $\Lambda$CDM and TDE are identical above $\ell \sim 30$. The high $\ell$ agreement implies the best-fit $\Lambda$CDM and TDE models have the same values of $\theta_s$ and the diffusion damping scale, $k_D$, which further implies matching values for $D_H(z_*)$ and $D_A(z_*)$ for the best-fit $\Lambda$CDM and TDE models \cite{Joudaki:2017zhq}. Thus, we can conclude that using $\Lambda$CDM-derived values for $D_H(z_*)$ and $D_A(z_*)$ for TDE inferences and GP inferences is self-consistent. It is likely the low $\ell$ disagreement arises from the late-time integrated Sachs-Wolfe effect and bears future study.

While the forecast TDE fitting and GP regression agree well on the preference for a transition in the dark energy evolution, the two methods show some differences in the details of the evolution. The GP inference allows for negative dark energy and so it has greater flexibility to fit both high and low-redshift data. By contrast, the dark energy density in the TDE model is constrained to be positive.
Thus, in order to fit the CMB's $D_A$, the fit favors a transition in $D_H$ that is sharper and at lower redshift than the transition in the median of the GP inference (see Figs.~\ref{fig:main} and \ref{fig:MCMC_dist_ztdzw0w1}).

Using \texttt{CLASS}, we calculate the various measures of the growth of perturbations, such as the matter power spectra and $\sigma(R,z)$. We also compare the growth function from \texttt{CLASS} with the less model-dependent solution to Eq.~\ref{eq:growth} in the Appendix. The TDE model can change $\sigma_8$ through the clustering of dark energy (depending on the microphysics) and the change in the growth function due to the modified expansion history~\cite{Fang:2008sn}. We assume that the dark energy density does not cluster significantly and, in keeping with this assumption, we keep the primordial power spectrum and transfer function fixed to that in $\Lambda$CDM but calculate the growth function at late times from our TDE dark energy model using {\tt CLASS}.

The resulting inferences of $f\sigma_8$ and $\sigma_8$ at $z=\{0, 0.5, 2\}$ are shown in Fig.~\ref{fig:MCMC_H0_de_growth}. Relative to the $\Lambda$CDM expectation, we find a noticeably slower growth rate today and at $z=0.5$ ($\Delta f \sigma_8 \simeq 0.05$ for both redshifts),  and mildly larger at $z=2$ (by approximately 1--2$\sigma$). We also find that $\sigma_8$ at present is smaller in the preferred TDE model, at mild significance ($\simeq2\sigma$). The predictions from the TDE model are consistent with the current measurements of $f\sigma_8$ at $z \lesssim 2$~\cite{Aghanim:2018eyx}. The differences from the $\Lambda$CDM predictions for $f\sigma_8$ are small compared to the uncertainties in current growth rate measurements but measurable by future surveys~\cite{surveywfirst, surveyhetdex, surveyeuclid, surveydesi, survey4most, surveyeboss, surveypfs}.

\section{SZ cluster abundance}\label{sec:SZ}

As a concrete test of the observable differences between the TDE model and $\Lambda$CDM, we focus on the SZ cluster abundance. Using $\sigma(R,z)$ for the $\Lambda$CDM and TDE models, we can calculate the expected number density per unit mass of gravitationally collapsed objects,
\begin{equation}
    \frac{dN}{dM dV}(M,z) = -\frac{\rho}{M^2} f_m(\sigma(M,z)) \frac{d\ln\sigma}{d \ln M},
\end{equation}
where $N$ is the number of clusters in some volume $V$, $M$ is the mass of the clusters, and $\rho$ is the matter density.
The multiplicity function $f_m(\sigma(M,z))$ is determined by fitting $dN/dV/dM$ to large volume $N$-body simulations, and we use the fitting function from Tinker~et~al.~(2008)~\cite{Tinker:2008ff}.

The number of clusters per unit mass per redshift is
\begin{equation}
    \frac{dN}{dMdz}= \frac{dN}{dMdV}\frac{dV}{dz},
\end{equation}
where $dV/dz =4\pi D_C^2 D_H$. We evaluate these functions at masses that for the two cosmologies give the same values of the SZ flux $Y_{500}$. Using the $M_{500} - Y_{500}$ relation from~\cite{Ade:2015fva}, we use $5\times10^{14} \rm M_\odot$ for the $\Lambda$CDM case and $4.5\times10^{14} \rm M_\odot$ (the redshift dependence of the relation is averaged over the redshift window where the SZ clusters are observed $0<z<0.4$) for the TDE case. The results of this calculation are shown in Fig.~\ref{fig:tanh}. There are two main sources of differences in the expected number of clusters in a redshift survey between $\Lambda$CDM and TDE models. One is in the multiplicity function through the dependence of $\sigma(R,z)$ on the TDE parameters and the other is in $dV/dz$ through the distances. We find that $f_m(\sigma(M,z))$ is smaller for the TDE model by 2--10\% between $z=1$ and $z=0$. For the volumetric factor, the TDE model predicts a roughly 15--30\% reduction in cluster counts between $z=1.5$ and $z=0$.  Together, the smaller volumetric factor and smaller growth factors work to suppress the number of clusters relative to $\Lambda$CDM by 15--40\% at these redshifts.

For tomographic cosmic shear measurements, similar considerations will apply and we expect the angular power spectra to be suppressed. We leave the potential of weak lensing and redshift space distortions to test these models for future work.

\section{Internal consistency of low-z distance measurements}\label{sec:bao}

\begin{figure}
    \centering
    \includegraphics[width=0.24\textwidth]{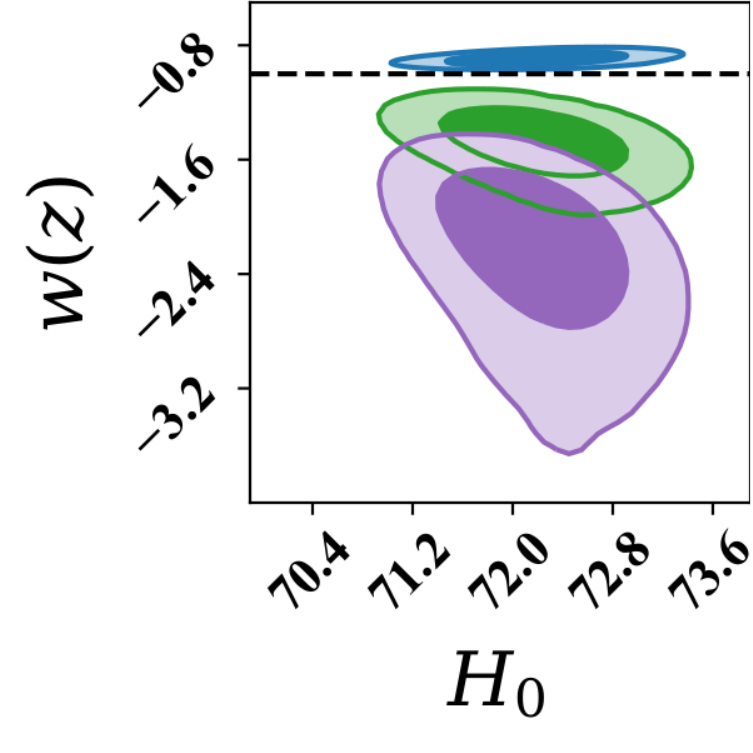}
    \includegraphics[width=0.24\textwidth]{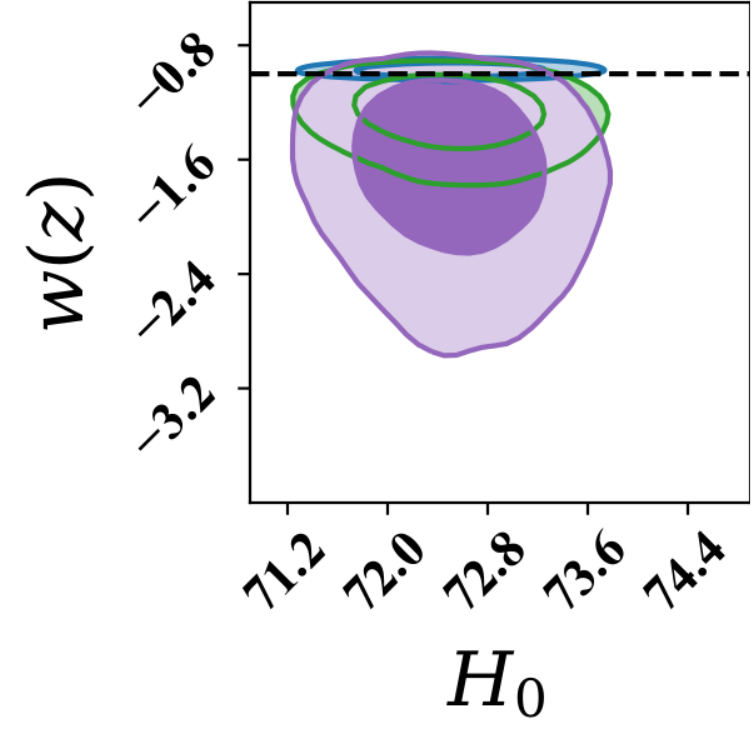}
    \caption{Posteriors of $H_0$ and $w(z)$ for $z=0,0.5,2.0$ (blue, green violet) for the cases where $r_{\rm drag}$ is varied independently (left) and scaled linearly with $D_H(z_*)$ (right). The black dashed line corresponds to the $\Lambda$CDM equation of state $w=-1$.}
    \label{fig:vardrag_w}
\end{figure}

Recent work \cite{Aylor:2018drw} has highlighted the tension between the BAO distances, calibrated to the value of $r_{\rm drag}$ from the $\Lambda$CDM fit to the Planck data, and the SN distances, calibrated to $H_0$. This tension is also present in our analysis. A possible resolution \cite{Aylor:2018drw} is that $r_{\rm drag}$ is smaller than the value inferred for $\Lambda$CDM from Planck, but here we have assumed that there is no new physics at $z>1000$ and hence $r_{\rm drag}$ is unchanged.

In this section, we report on results when deviating from our main analysis in two ways.  Fig.~\ref{fig:vardrag_w} summarizes these results.
In the first test, we allow $r_{\rm drag}$ to vary independently of other distances and scale BAO distances accordingly. A transition in the DE density between $z=2$ and today is still inferred, but the equation of state varies more gently. The recovered value of $r_{\rm drag}$ is smaller, indicating the presence of a real tension between the BAO and SNe + $H_0$ datasets \cite{Aylor:2018drw}. We discuss these points in greater detail with relevant plots in the Appendix.

In the second test, we allow $r_{\rm drag} \propto D_H(z_*)$ as an illustrative example to explore the degeneracy between new physics at early ($z>1000$) and late times ($z<3$). We allow the errors on $D_H(z_*)$ and $D_A(z_*)$ to be larger (TT+TE+EE+lowP constraints on $\Lambda{\rm CDM} + N_{\rm eff}$ model in J18),
as would be expected with the addition of new parameters. The inferred errors on the dark energy evolution are larger and it is not possible to reach a strong conclusion about the DE density at $z=2$, although a sharp transition in the TDE equation of state is still allowed.
\vspace{4.5mm}

\section{Conclusions}\label{sec:discussion}

We performed a GP regression for the expansion history of the Universe using Planck measurements of the CMB, BOSS measurements of the BAO signal in the Ly$\alpha$ forest and LRGs, Pantheon compilation of Type Ia SNe, and a measurement of the present Hubble parameter with forecasted 1\% uncertainty. The forecast regression prefers a dark energy component with equation of state $w>-1$ at present, and has the density transition to zero by $z\simeq2$. An interesting corollary of our result is the wide range of possibilities for the equation of state in the future, with a de Sitter phase not being favored. We calculated the growth history assuming no extra sources of clustering except for matter, and showed that the inferred growth rate in this model is measurably different from the Planck $\Lambda$CDM expectation.

Our forecast GP results are recovered when using a parametric model for dark energy evolution that allows for a sharp transition in the dark energy density. We used \texttt{CLASS} to calculate the predictions of this TDE model for SZ cluster counts and found that the TDE model predicts noticeably less SZ clusters than the best-fit $\Lambda$CDM model, potentially alleviating the $\sigma_8$ tension.

Similar, but less sharp, results were found when we allowed $r_{\rm drag}$ to vary independently of the CMB distances to explore the internal consistency between the $z<1$ distance measurements.  However, when $r_{\rm drag}$ was taken to scale linearly with $D_H(z_*)$ as an illustrative example of new physics at $z>1000$, an evolving dark energy component was still allowed but not strongly preferred.  In this case, the low-redshift distances agree better but that comes at the cost of not fitting the CMB precisely.

Direct reconstruction of the Universe's expansion history via the BAO signal that will be observed by future surveys, such as DESI, LSST, WFIRST, and Euclid should be able to robustly detect a transition in the dark energy equation of state. The fact that the TDE model predicts less growth of perturbations than $\Lambda$CDM offers another way to test this model through redshift space distortion measurements and tomographic weak lensing analysis in the future.

Our results suggest that a sharp transition in the dark energy equation of state for $1<z<2$ could simultaneously explain the $H_0$ and $\sigma_8$ tensions.

\section{Acknowledgments}
We are grateful to Lloyd Knox for discussions that prompted us to focus on the internal consistency of low redshift distance measurements. MK was supported by National Science Foundation Grant No. PHY-1620638.
SJ acknowledges support from the Beecroft Trust and ERC 693024. This work was supported by the high performance computing cluster Seondeok at the Korea Astronomy and Space Science Institute.

\bibliographystyle{plain}
\bibliography{sample}

\appendix

\section{Growth rate calculation}\label{sec:growthcheck}
\begin{figure}
    \centering
    \includegraphics[width=0.45\textwidth]{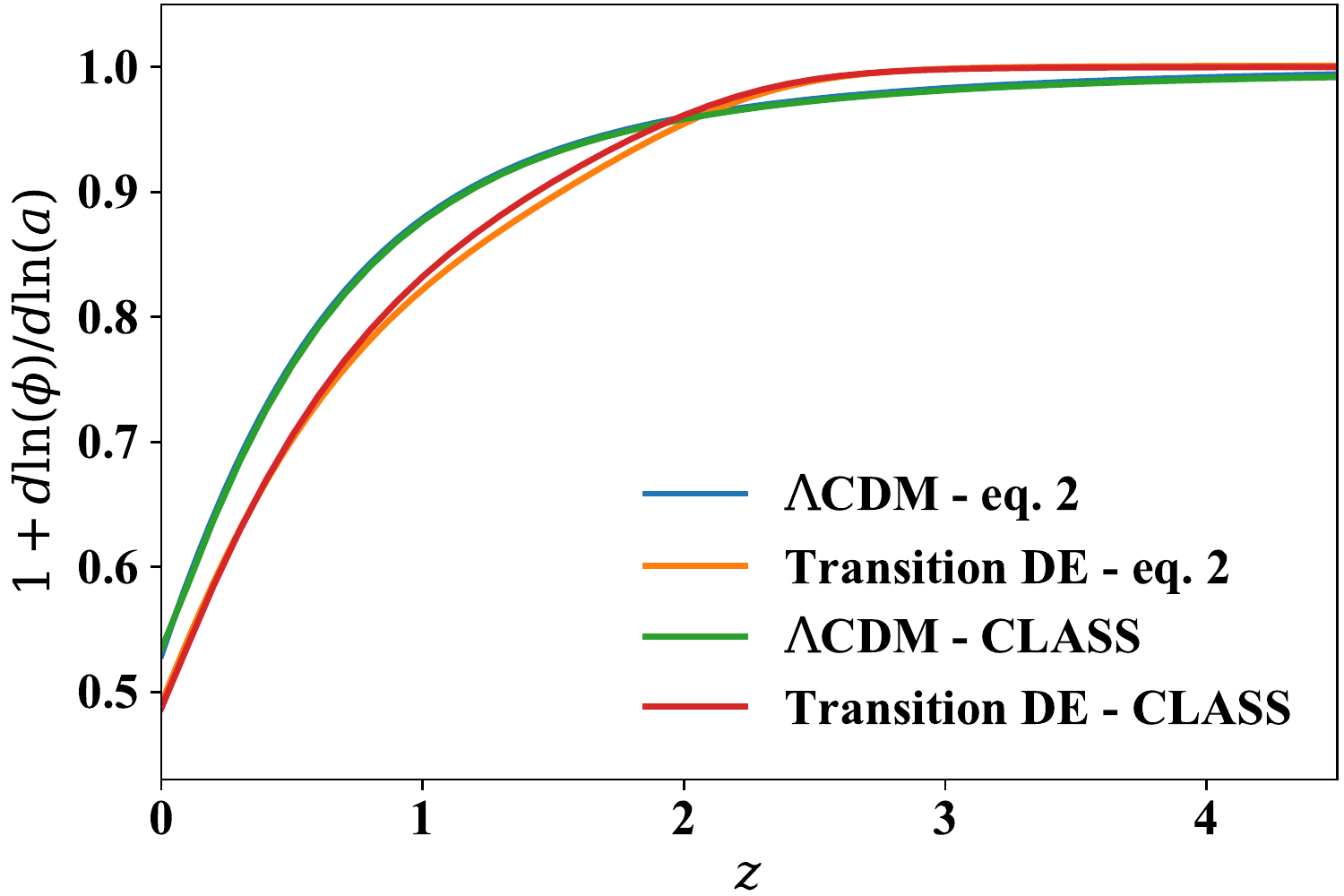}
    \caption{The growth rates for $\Lambda$CDM and TDE expansion histories, as calculated by Eq.~\ref{eq:growth} and as calculated by \texttt{CLASS}. The results from Eq.~\ref{eq:growth} for $\Lambda$CDM and TDE are in blue and orange, respectively. The results from \texttt{CLASS} for $\Lambda$CDM and TDE are in green and red, respectively.
    }
    \label{fig:growth_check}
\end{figure}

Here we examine the consistency between the growth rate calculated from Eq.~\ref{eq:growth} and the growth rate extracted from \texttt{CLASS}. Specifically, we use the TDE parameters that match the median GP inference ($w_0=-0.95$, $w_1=-1.95$, $z_t=2.5$, and $\Delta_z=0.9$). Fig.~\ref{fig:growth_check} shows that in $\Lambda$CDM there is effectively no difference between the two methods to calculate the growth rate. For our TDE dark energy model there is a small but noticeable deviation at intermediate redshifts, in other words, at redshifts where the dark energy equation of state is varying rapidly. However, these deviations are much smaller than the difference from the $\Lambda$CDM growth rate.

\section{Transitional Dark Energy model and the CMB}\label{app:CMBTDE}

\begin{figure}
    \centering
    \includegraphics[width=0.45\textwidth]{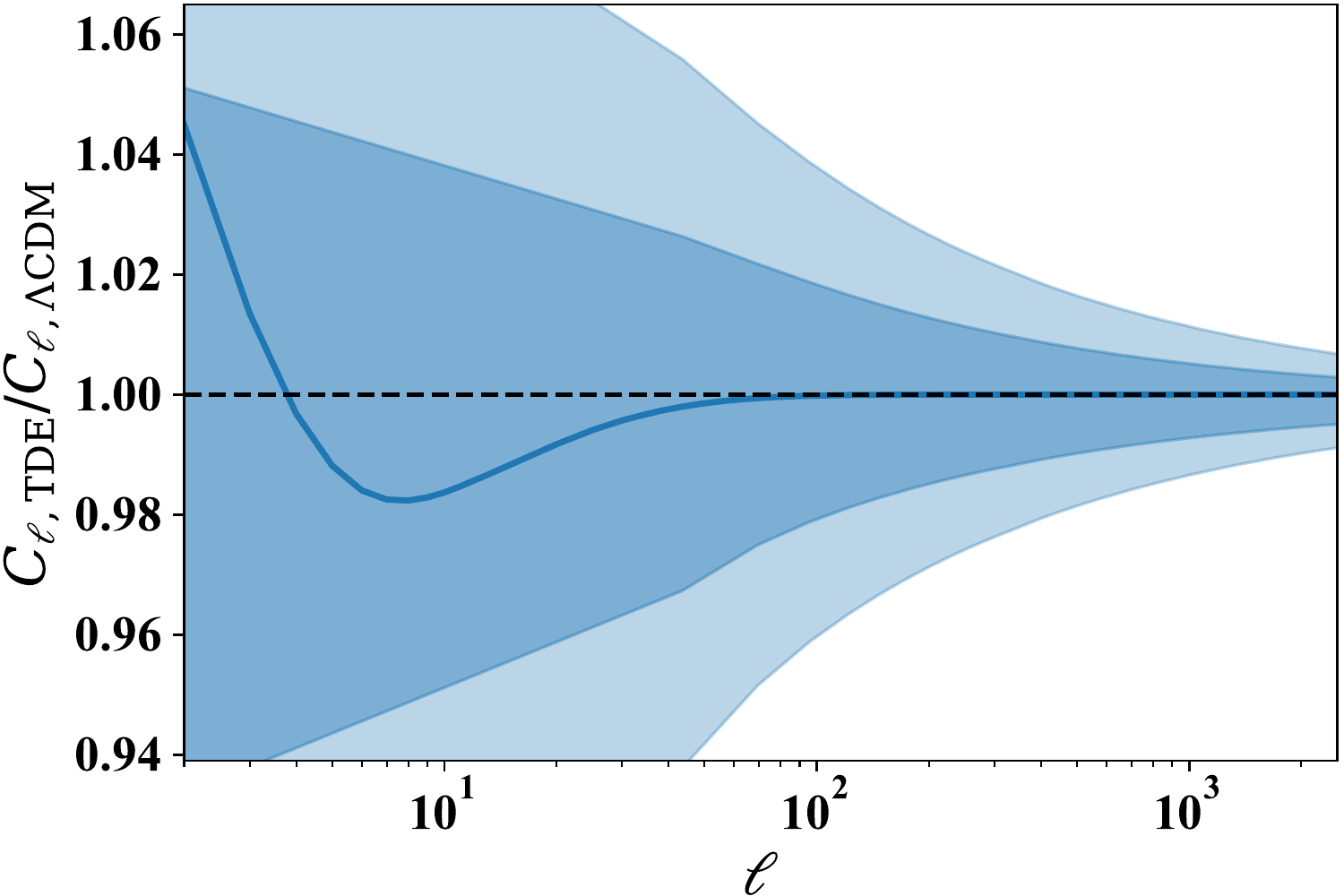}
    \caption{Ratio of $C_\ell$ for the temperature power spectra for a TDE model with best-fit parameters from the MCMC run
    (such that it predicts the $\Lambda$CDM values for $D_H(z_*)$ and $D_A(z_*)$) relative to the $C_\ell$ for the fiducial $\Lambda$CDM model.  The bands correspond to the cosmic variance 68\% and 95\% CLs.
    }
    \label{fig:Cell}
\end{figure}

\begin{figure*}
    \centering
    \includegraphics[width=\textwidth]{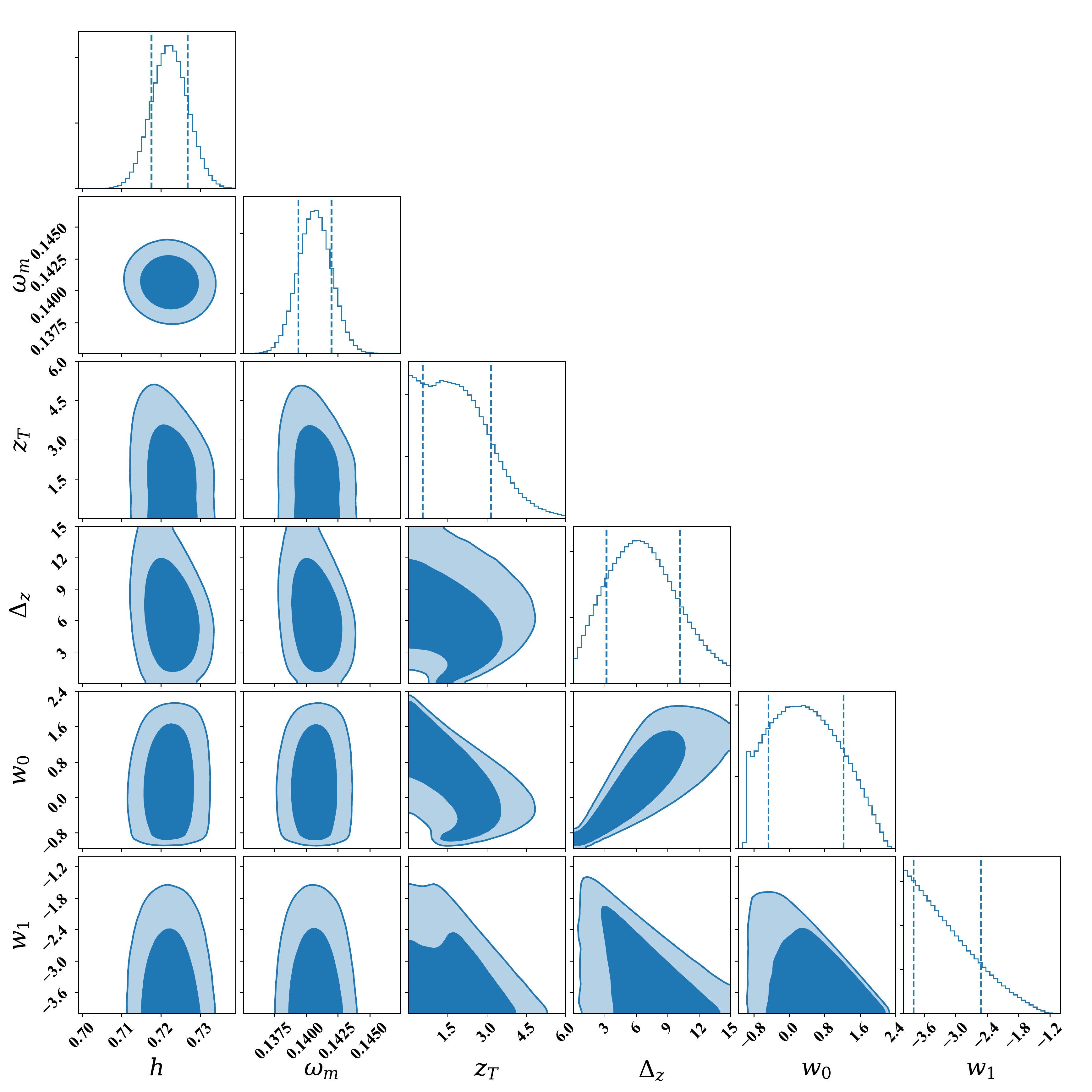}
    \caption{Posterior for an MCMC inference of the TDE model where $z_T$, $\Delta_z$, $w_0$, and $w_1$ are varied, along with the $\Lambda$CDM parameters $h$ and $\omega_{\rm m}$. The full dataset (CMB+Ly$\alpha$+LRG+SN+$H_0$) from our GP analysis is being fit. Note the $H_0$ constraint here is the 1\% projected error. For the 2D posteriors, the 68\% and 95\% CLs are shown in increasingly lighter shades of blue. For the 1D posteriors, the 68\% CL is encapsulated by the dashed vertical lines. It is informative that this model can fit $h$ and $\omega_{\rm m}$, the parameters that characterize both the low and high redshift anchors of the expansion history.  It is less clear how to interpret the parameters that characterize the transition in the dark energy equation of state. Therefore we plot $\rho_{DE}(z)$ and $w(z)$ at various redshifts in Figs.~\ref{fig:MCMC_H0_de_growth} and \ref{fig:MCMC_derived} to show how these parameters work physically.}
    \label{fig:full_triangle}
\end{figure*}

\begin{figure*}
    \centering
    \includegraphics[width=\textwidth]{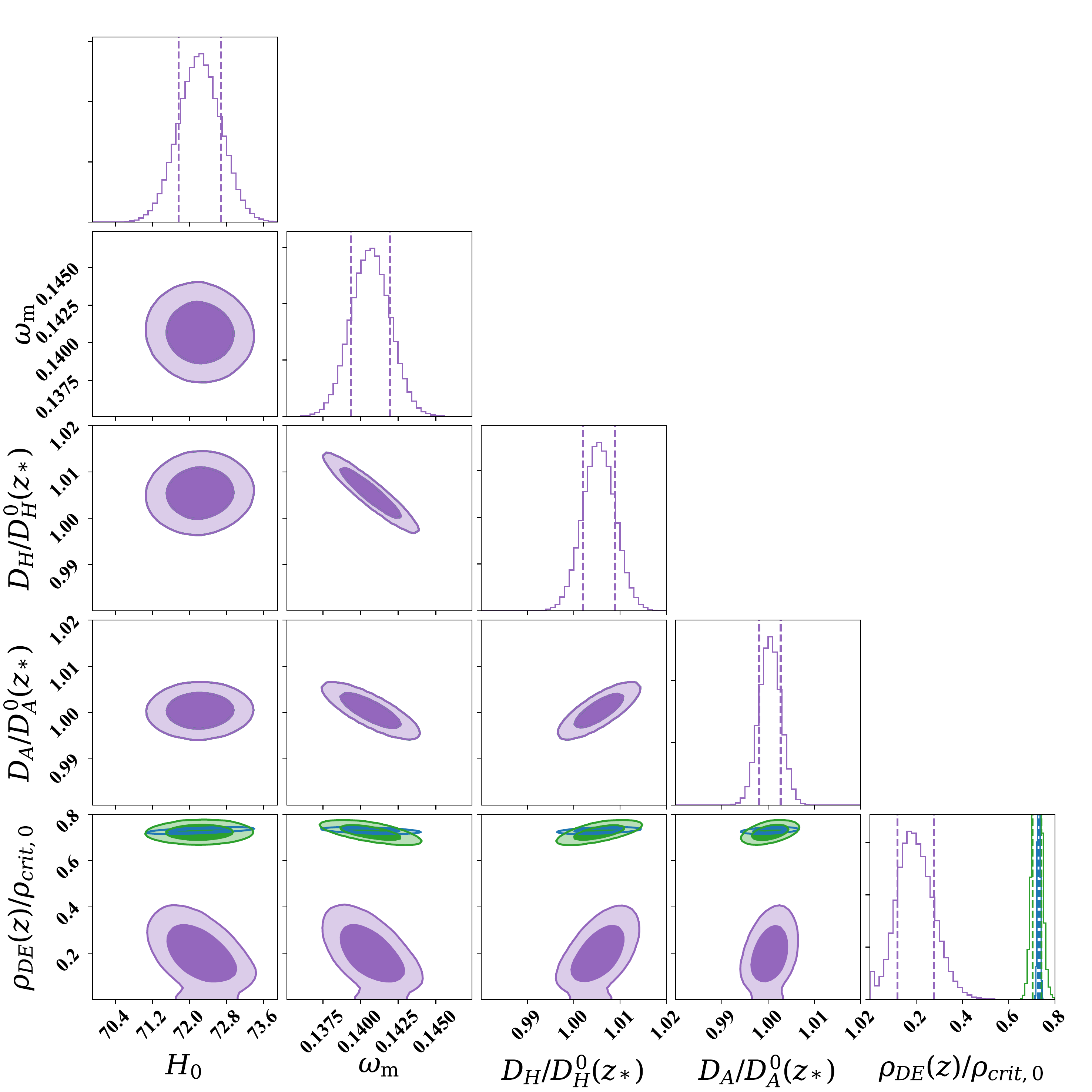}
    \caption{Same case as Fig.~\ref{fig:full_triangle}, except  instead of the TDE parameters the quantities being displayed are the dark energy density at redshifts $z=0,0.5,2.0$ (blue, green, violet) relative to the present value of the critical density, as well as $D_H(z_*)$ and $D_A(z_*)$ relative to the fiducial Planck 2015 values. These derived parameters are useful to show how inferences of the TDE model yield the same high redshift distances as $\Lambda$CDM, and to show how the dark energy density evolves physically.
    }
    \label{fig:MCMC_derived}
\end{figure*}

\begin{figure*}
    \centering
    \includegraphics[width=\textwidth]{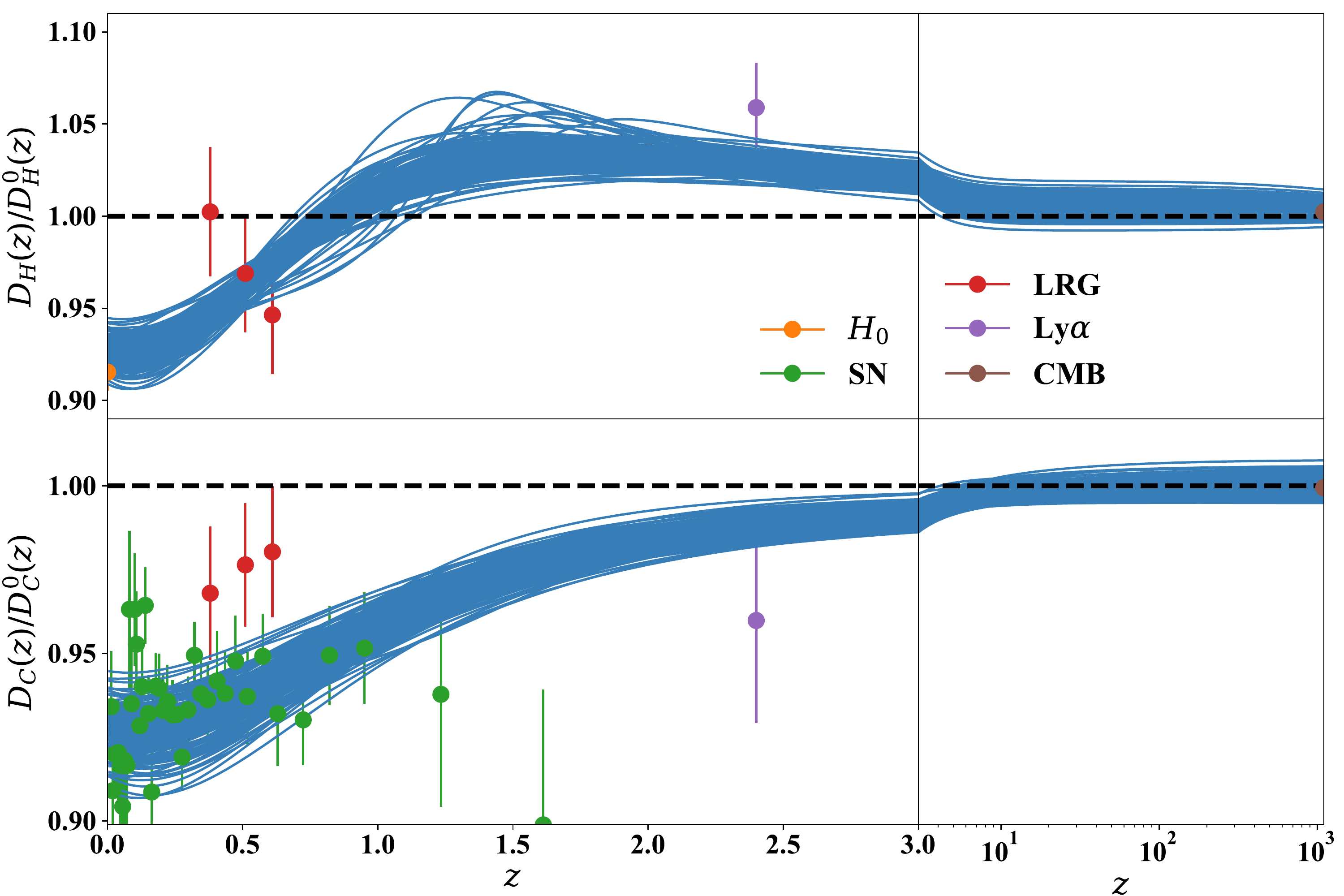}
    \caption{Posterior predictive distribution composed of sample expansion histories (blue) drawn from an MCMC sampling of the posterior of the TDE model as in Fig.~\ref{fig:full_triangle}.  The spread in these expansion histories corresponds to the posterior uncertainty in the expansion history. The data are plotted in various colors, orange for $H_0$, green for the SNe data, red for the LRG data, violet for the Ly$\alpha$ data, and brown for the CMB data.  The SNe constrain only the ratio of the distances $D_L / D_{H_0}$ so the absolute scale of the SNe data points in this figure is set by the best fit $H_0$.  Similarly, the BAO data points (LRG + Ly$\alpha$) only constrain the ratios $D_H/r_{\rm drag}, D_A/r_{\rm drag}$ so the absolute scale for these distances is set by the best fit value from Planck 2015 \cite{planck15}, $r_{\rm drag} = 147$.  These distances draw from the posterior are useful to demonstrate which features of the data are driving the need for a transition in the dark energy equation of state, primarily $D_A(z_*)$.
    }
\label{fig:MCMC_dist_ztdzw0w1}
\end{figure*}

Here we investigate to what extent our proposed TDE model modifies the anisotropies in the CMB. To this end, we used \texttt{CLASS} to calculate the  angular temperature power spectrum for both the fiducial $\Lambda$CDM cosmology and a TDE model with best-fit parameters from a fit to the full Planck 2018 likelihoods~\cite{Aghanim:2018eyx}. As seen in Fig.~\ref{fig:Cell}, the $C_{\ell}$s of the best-fit parameters of these two models agree exactly for $\ell \gtrsim 30$. This implies the best-fit $\Lambda$CDM and TDE parameters yield the same values for $\theta_s$ and $k_D$.  Hence, the best-fit TDE parameters yield $D_H(z_*) =0.190$ Mpc (which effectively sets $\omega_{\rm m}$) and $D_A(z_*)=12.6$ Mpc that match the $\Lambda$CDM values ($D_H(z_*)=0.190$ Mpc and $D_A(z_*)=12.6$ Mpc)~\cite{Joudaki:2017zhq}. Thus, it is self-consistent to use $\Lambda$CDM-derived values for TDE and GP inferences. In Figs.~\ref{fig:full_triangle},~\ref{fig:MCMC_derived} and \ref{fig:MCMC_dist_ztdzw0w1}, we show that the MCMC exploration of the TDE parameter space does result in the TDE model being tightly constrained around the $\Lambda$CDM values for $D_H(z_*)$ and $D_A(z_*)$. The discrepancies between TDE and $\Lambda$CDM $C_{\ell}$s at $\ell \lesssim 30$ is smaller than cosmic variance uncertainties and likely has implications for the late-time Sachs-Wolfe effect. We conclude that when the TDE model matches the $\Lambda$CDM values for $D_H(z_*)$ and $D_A(z_*)$, the $C_{\ell}$ cannot distinguish between the $\Lambda$CDM and TDE models.

\section{Scaling the BAO distances}

In this section of the Appendix, we explore the effects of allowing $r_{\rm drag}$ to vary on our inferences of the TDE model. The BAO distances scale with $r_{\rm drag}$ and if the value of $r_{\rm drag}$ that is preferred by a $\Lambda$CDM fit to the CMB is used for this scaling, the BAO distances disagree with the SN distances calibrated by the $H_0$ measurement \cite{Aylor:2018drw}, which is evident in Fig.~\ref{fig:MCMC_dist_ztdzw0w1}. The TDE model alone cannot explain this discrepancy and only lowers its significance. Therefore, we investigate cosmologies where $r_{\rm drag}$ is allowed to vary.

\begin{figure*}
    \centering
    \includegraphics[width=\textwidth]{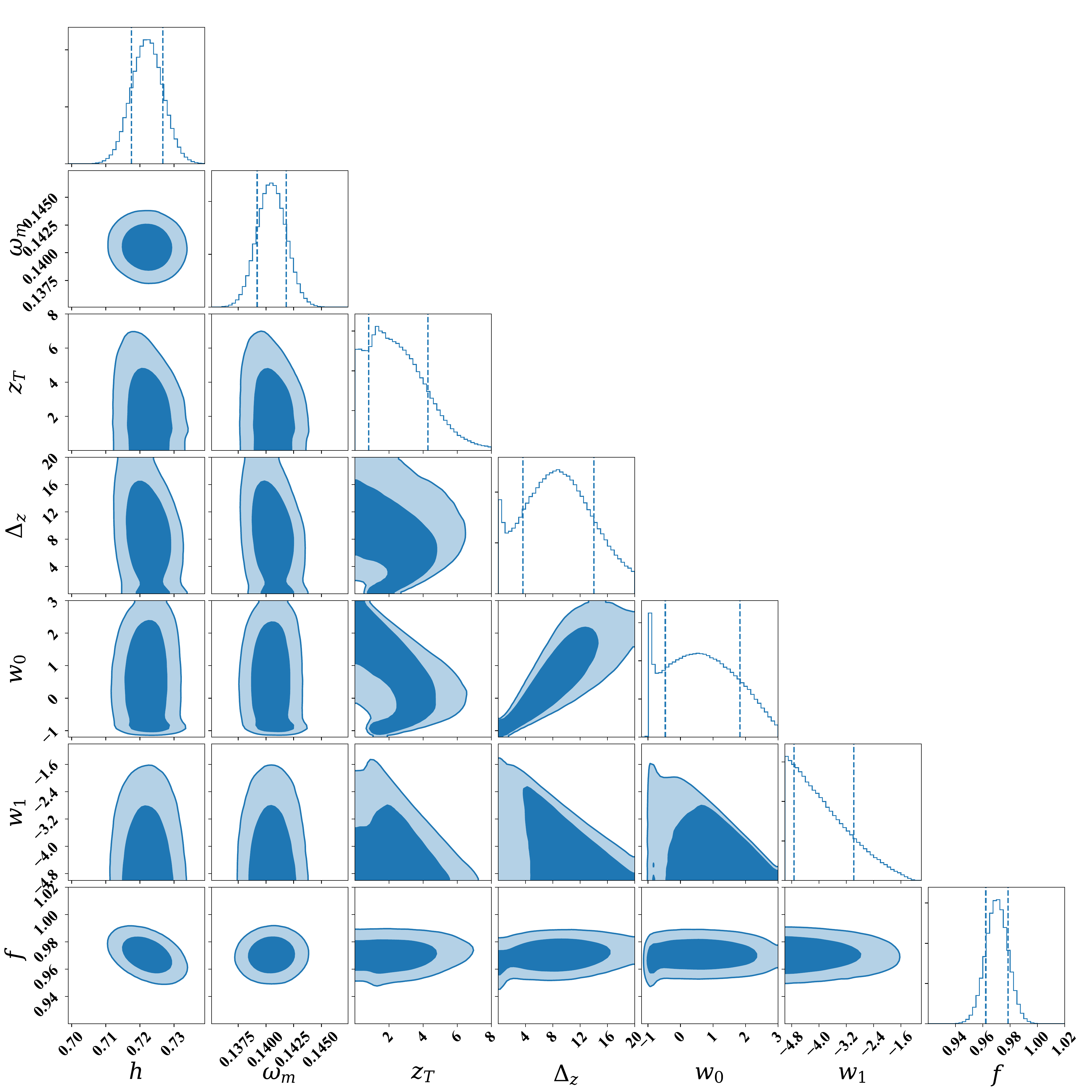}
    \caption{Posteriors for the TDE model but now $r_{\rm drag}$ is scaled by an extra parameter $f$. The fact $f$ is noticeably below 1 shows the magnitude of the disagreement between the Planck-$\Lambda$CDM calibrated LRG distances and the $H_0$ calibrated SNe distances.
    }
    \label{fig:fudge_params}
\end{figure*}

\begin{figure*}
    \centering
    \includegraphics[width=\textwidth]{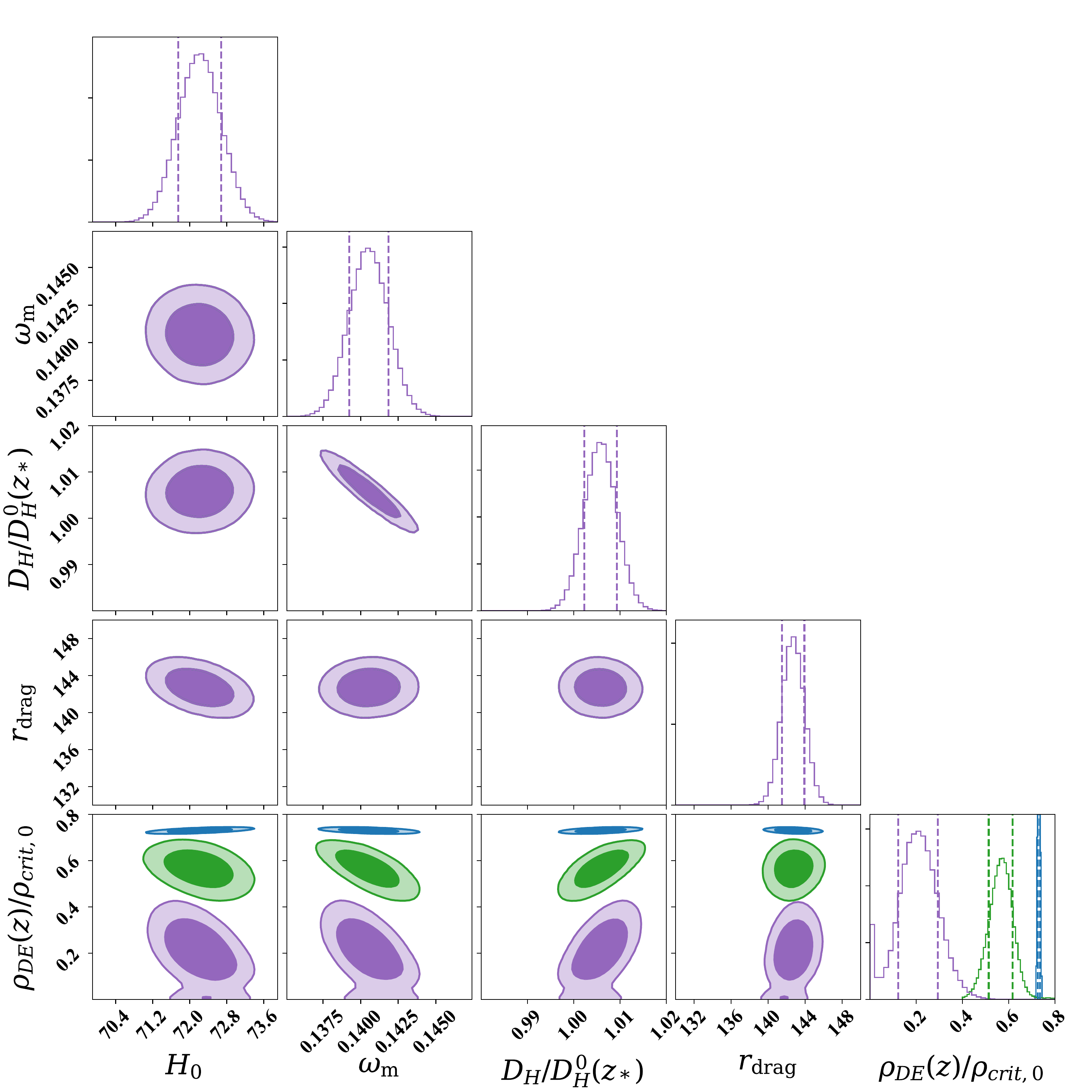}
    \caption{The same case as in Fig.~\ref{fig:fudge_params} but shown are the posteriors for the derived parameters including $D_H(z_*)$, $r_{\rm drag}$, and $\rho_{\rm DE}(z)/\rho_{\rm crit,0}$, for redshifts $z=0,0.5,2$ (blue, green, violet). As with Fig.~\ref{fig:MCMC_derived}, this figure shows how inferences of the TDE model yield the same values of $D_H(z_*)$ as $\Lambda$CDM, and shows how the dark energy density evolves physically. The $r_{\rm drag}$ posterior shows how it must be shifted to lower values than the $\Lambda$CDM value to get the LRG and SNe distances to agree.}
    \label{fig:fudge_derived}
\end{figure*}

\begin{figure*}
    \centering
    \includegraphics[width=\textwidth]{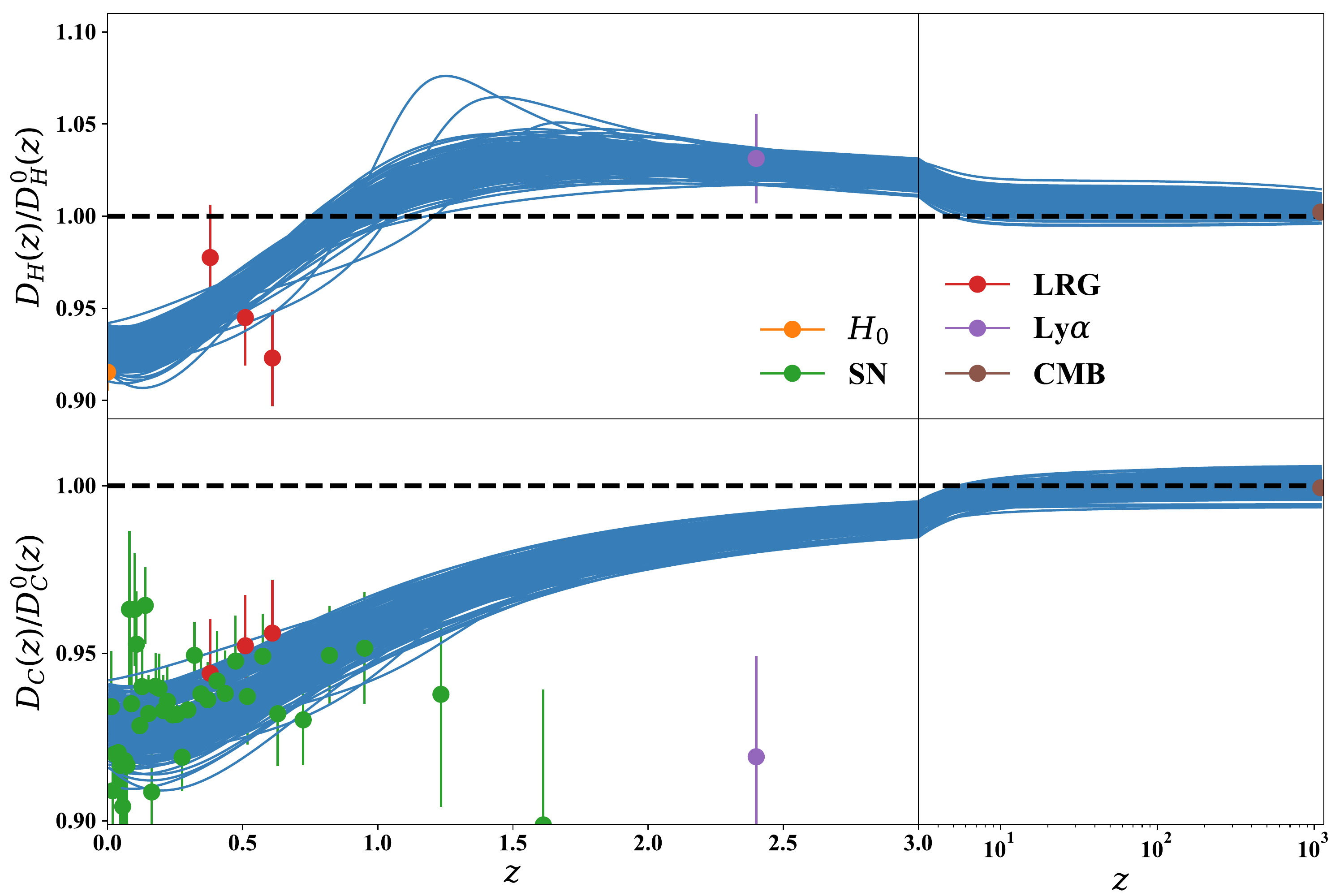}
    \caption{The same case as in Fig.~\ref{fig:fudge_params} but shown is the posterior predictive distribution, as in Fig.~\ref{fig:MCMC_dist_ztdzw0w1}. The absolute scale of the SNe data points are again set by the best fit value of $H_0$ while the absolute scale for the BAO data points are set by the best fit $r_{\rm drag}$, which in this case, is allowed to vary freely. As with Fig.~\ref{fig:MCMC_dist_ztdzw0w1}, these posterior-sampled distances are useful to demonstrate which features of the data are driving the need for a transition in the dark energy equation of state, primarily $D_A(z_*)$. The difference with Fig.~\ref{fig:MCMC_dist_ztdzw0w1} is that now $r_{\rm drag}$ is allowed to scale freely so the LRG and SNe distances agree.}
    \label{fig:fudge_dist}
\end{figure*}

We explore two cases, one where $r_{\rm drag}$ is allowed to vary independently of any other parameter, and another where it scales linearly with $D_H(z_*)$. The results for the first case are shown in Figs.~\ref{fig:fudge_params}--\ref{fig:fudge_dist}.  The most notable features of this case are that allowing $r_{\rm drag}$ to vary independently causes the low-redshift distances to completely agree (as expected), and also still shows a preference for little to no dark energy density at redshift $z=2$. The evolution of the dark energy density between redshifts $z=0$ and $z=2$, however, is more gradual than the case where $r_{\rm drag}$ is fixed.

The Bayes factor for this model, relative to the $\Lambda$CDM model, is $\ln K = 7.9$ in favor of the TDE+$r_{\rm drag}$ model, or, relative to the TDE model, is $\ln K =1.1$.
This indicates no substantial preference, and that allowing the BAO distances to vary freely, on top of the TDE model, is only mildly preferred.  Interestingly, there is moreover no preference for the $\Lambda$CDM+$r_{\rm drag}$ model over the $\Lambda$CDM-only model ($\ln K = -0.7$). This is because the preferred value of $h$ in this case is around 0.7, and so the SN distances already agree with the BAO distances, thus leaving no work for the freely varying $r_{\rm drag}$ to perform.

The $r_{\rm drag}$ value inferred is smaller than the $\Lambda$CDM prediction by approximately 3\%, which reflects the internal inconsistency between the $z<1$ distances~\cite{Aylor:2018drw}. Whether this is due to an unknown systematic in the BAO distances or a signal for new physics that is relevant at $z>1000$ is not evident.

\begin{figure*}
    \centering
    \includegraphics[width=\textwidth]{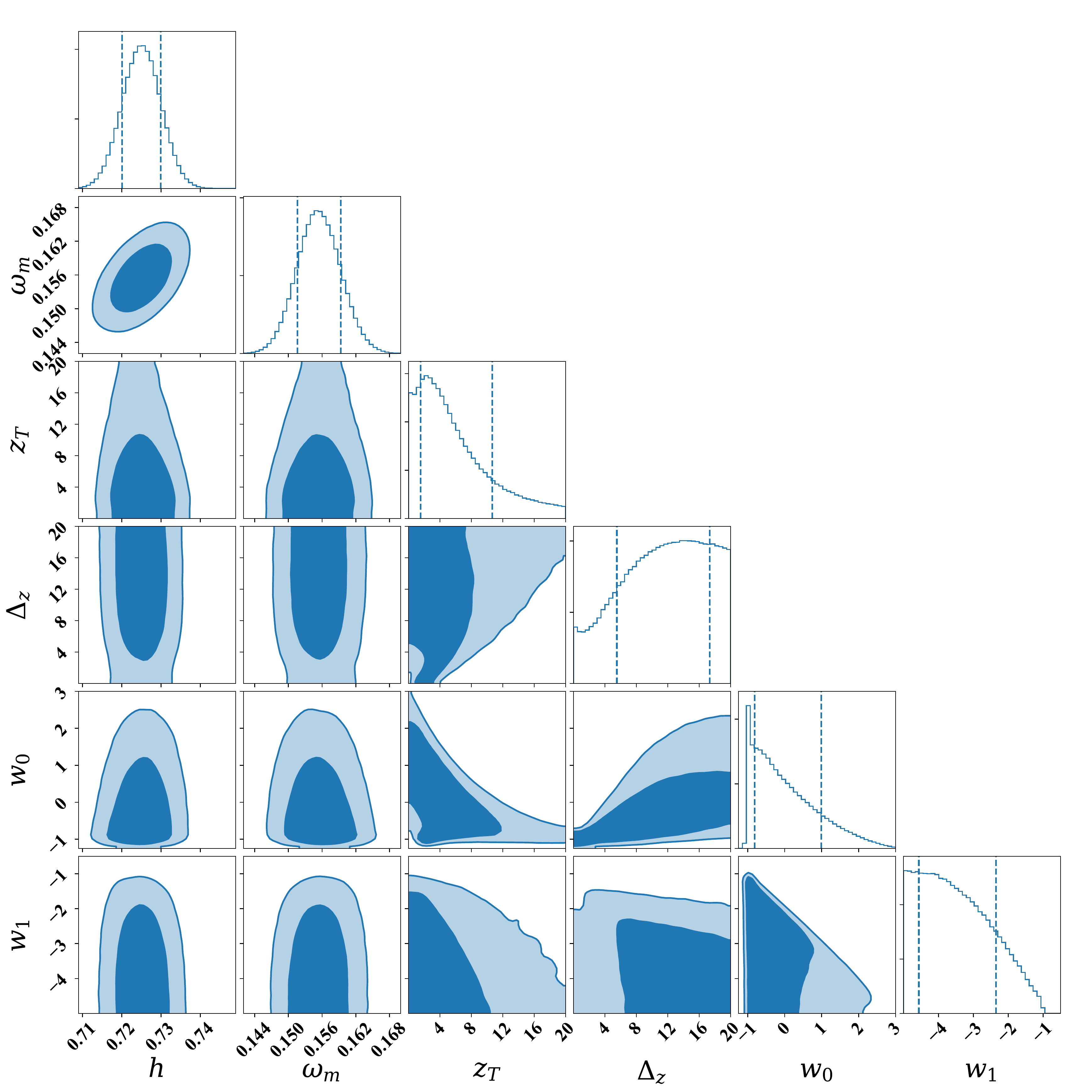}
    \caption{Posteriors for the TDE model but now $r_{\rm drag}$ scales linearly with $D_H$.  The derived values of $\omega_m$ are noticeably different than in Figs.~\ref{fig:full_triangle} and \ref{fig:fudge_params} since the need to modify $r_{\rm drag}$ requires a corresponding change in $\omega_m$.
    }
    \label{fig:linscale_params}
\end{figure*}

\begin{figure*}
    \centering
    \includegraphics[width=\textwidth]{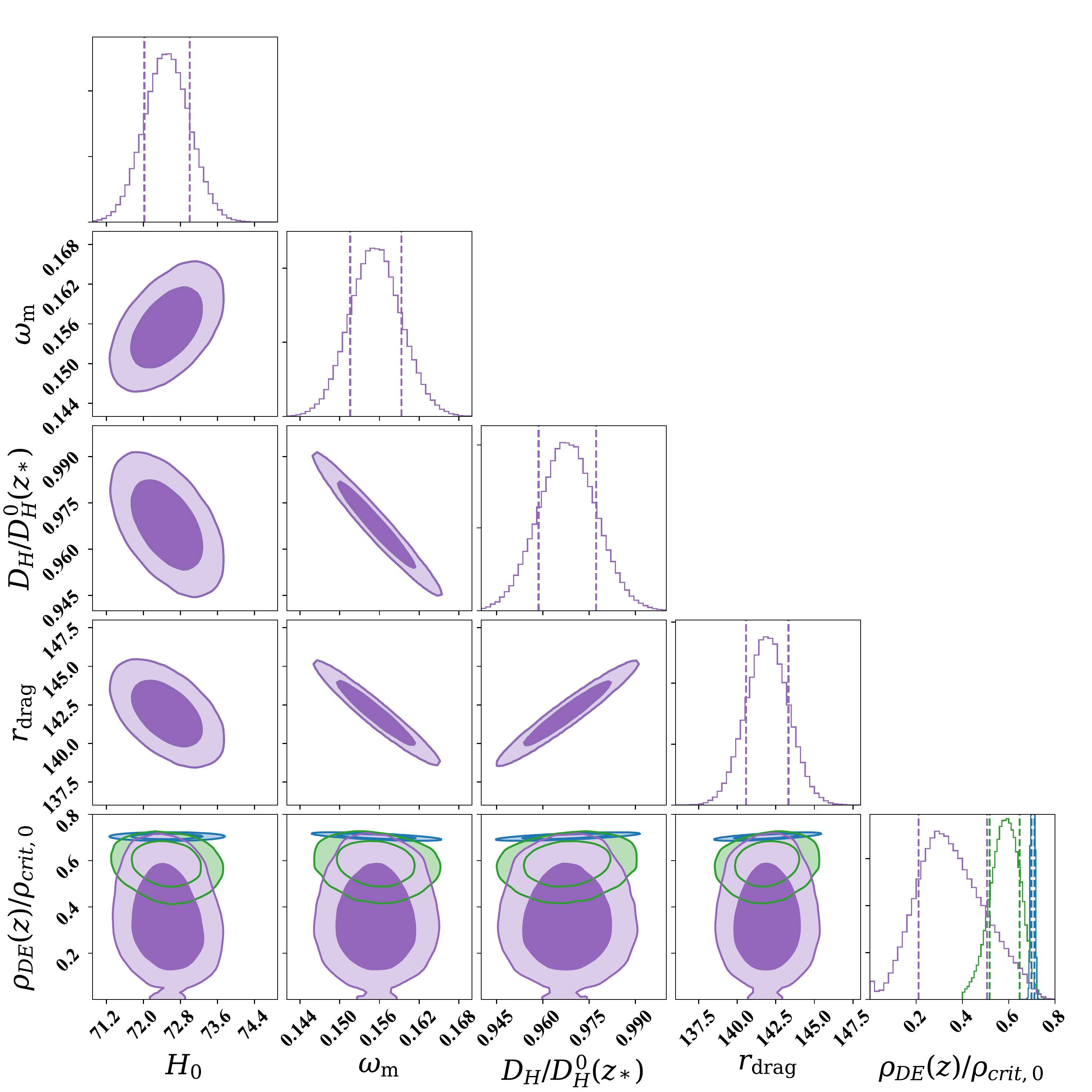}
    \caption{The same case as in Fig.~\ref{fig:linscale_params} but shown are the posteriors for the derived parameters including $D_H(z_*)$, $r_{\rm drag}$, and $\rho_{\rm DE}(z)/\rho_{\rm crit,0}$, for redshifts $z=0,0.5,2$ (blue, green, violet).  Similar to Figs.~\ref{fig:MCMC_derived} and \ref{fig:fudge_derived}, this plots is useful to show how the TDE model with $r_{\rm drag}$ scaling linearly with $D_H(z_*)$ predicts values for $\omega_m$, $D_H(z_*)$ and $r_{\rm drag}$ different than the $\Lambda$CDM values, while showing no strong preference for a transition in the dark energy density.}
    \label{fig:linscale_derived}
\end{figure*}

\begin{figure*}
    \centering
    \includegraphics[width=\textwidth]{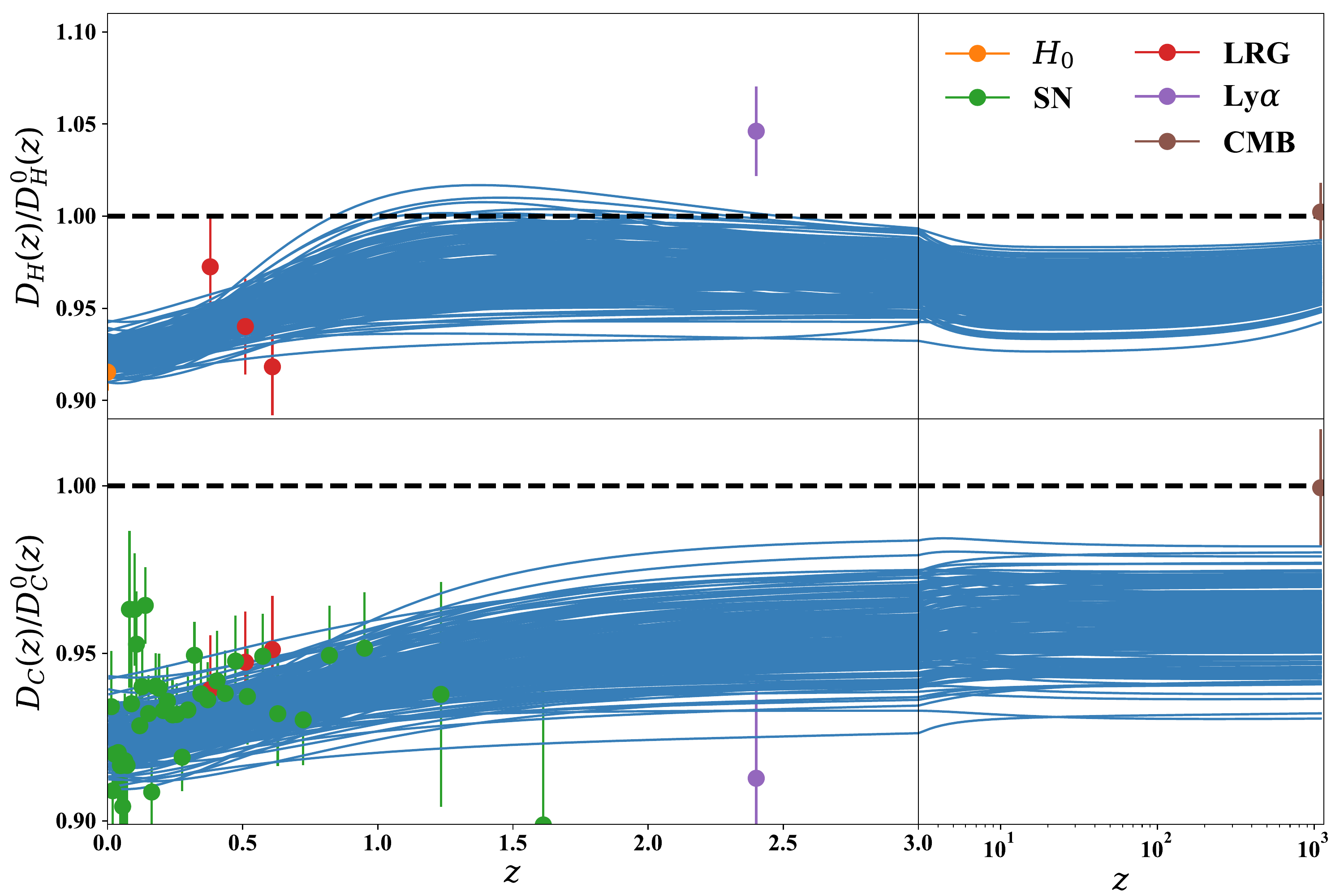}
    \caption{The same case as in Fig.~\ref{fig:linscale_params} but shown is the posterior predictive distribution, as in Fig.~\ref{fig:MCMC_dist_ztdzw0w1}. The absolute scale of the SNe are determined as in Fig.~\ref{fig:MCMC_dist_ztdzw0w1}, while the absolute scale for the BAO data points are set by the best fit $r_{\rm drag}$, which in this case, varies linearly with $D_H(z_*)$.  As with Figs.~\ref{fig:MCMC_dist_ztdzw0w1} and \ref{fig:fudge_dist}, this plot is useful in showing what features of the data are driving the results.  The larger uncertainties on the CMB distances, combined with the preference to shift the LRG distances down to the SNe distances, shifts the whole expansion history down, rather than invoke a transitional dark energy density.
    }
    \label{fig:linscale_dist}
\end{figure*}

When $r_{\rm drag}$ is tied to $D_H(z_*)$, a somewhat different picture emerges, as shown in Figs.~\ref{fig:linscale_params}--\ref{fig:linscale_dist}.  This scaling relationship is supposed to represent a more physically motivated way to vary $r_{\rm drag}$. Such a scaling would arise if some variation left the sound speed of the pre-recombination plasma unchanged while causing the expansion history to scale up or down. This could happen, for example, if there is extra radiation or early dark energy \cite{Aylor:2018drw,Bernal:2016gxb,Poulin:2018cxd}. To be self consistent, we also increased the error on $D_H(z_*)$ and $D_A(z_*)$ to be the same as that obtained from TT+TE+EE+lowP constraints in a $\Lambda$CDM + $N_{\rm eff}$ model (see Table I in J18). If the size of the errors on $D_H(z_*)$ and $D_A(z_*)$ are obtained from the $\Lambda$CDM case, then the same results emerge, whether the $r_{\rm drag}$ is tied to $D_H(z_*)$ or fixed. In the $\Lambda$CDM + $N_{\rm eff}$ case, the shifts in the distances relative to the $\Lambda$CDM values were not significant but the uncertainty increased by a factor of 3.

Using this more conservative constraint, we calculated the posterior for our TDE model where $r_{\rm drag}$ scales linearly with $D_H(z_*)$.  This scaling shifts the whole distance ladder down by $\sim$5\% relative to the fiducial Planck $\Lambda$CDM values (see Fig.~\ref{fig:linscale_dist}). No strong evidence for dark energy evolution was found, though such evolution remains allowed, as shown in Fig.~\ref{fig:linscale_derived}. The picture is essentially the same if $r_{\rm drag}$ scales as $D_H(z_*)^{1.5}$, which is the best-fit scaling relation calculated in J18 for the $\Lambda$CDM + $N_{\rm eff}$ model. The Bayes factor for the TDE model relative to $\Lambda$CDM, both with $r_{\rm drag}$ tied to $D_H(z_*)$, is $\ln K = -0.8$. This indicates no substantial preference between the TDE+$r_{\rm drag}$ and the $\Lambda$CDM+$r_{\rm drag}$ models.

The ability of the model where $r_{\rm drag}$ and $D_H(z_*)$ vary in tandem to alleviate the tensions in cosmological distances relies on the fact that the CMB constraint is less stringent. This can be seen in Fig.~\ref{fig:linscale_dist}, where the preferred high redshift distances are pulled away from the centers of the CMB constraint, beyond the $1\sigma$ range. If the uncertainties in the constraint were reduced to their size in $\Lambda$CDM, then any preference for the BAO distances to be scaled to the values picked out by the SNe disappears, and the strong preference for the TDE reappears.  Unless the BAO distances are disconnected from the CMB distances, there remains some degree of tension between the inferred distances, either at $z=0$ in the $\Lambda$CDM model, at $z\sim 0.5$ in the TDE model, or at $z=z_*$ for the $\Lambda$CDM (or TDE) model wherein $r_{\rm drag}$ and $D_H(z_*)$ covary.

\end{document}